\documentclass[%
superscriptaddress,
preprint,
 amsmath,
 amssymb,
 aps,
 pre,
]{revtex4-2}

\usepackage[english]{babel}
\usepackage{graphicx}
\usepackage{dcolumn}
\usepackage{bm}

\newcommand{\Pe}{\operatorname{Pe}}
\newcommand{\Cn}{\operatorname{Cn}}
\newcommand{\Ca}{\operatorname{Ca}}
\renewcommand{\Re}{\operatorname{Re}}
\newcommand{\deltayzf}{\ell}

\usepackage{gensymb}
\usepackage{subfig}

\DeclareMathSymbol{\shortminus}{\mathbin}{AMSa}{"39}

\usepackage{algorithmic}
\usepackage[lined,boxed,ruled,commentsnumbered]{algorithm2e}
\usepackage{xcolor}
 
\newcommand{\change}[1]{\textcolor{black}{#1}}

\begin{document}

\title{Mass diffusion and bending in dynamic wetting\\ by phase-field and sharp-interface models}%

\author{Tomas Fullana}
\affiliation{Institut Jean le Rond $\partial$'Alembert, Sorbonne Universit{\'e}, CNRS, F-75005 Paris, France}
\affiliation{Laboratory of Fluid Mechanics and Instabilities,
École Polytechnique Fédérale de Lausanne, Lausanne, CH-1015, Switzerland}
\author{Stéphane Zaleski}
\affiliation{Institut Jean le Rond $\partial$'Alembert, Sorbonne Universit{\'e}, CNRS, F-75005 Paris, France}
\affiliation{Institut Universitaire de France, Paris, France}
\author{Gustav Amberg}%
 \email{gustava@kth.se}
\affiliation{Flow Centre, Department of Engineering Mechanics, The Royal Institute of Technology, 100 44 Stockholm, Sweden}
\begin{abstract}
Dynamic wetting poses a well-known challenge in classical sharp interface formulation as the no-slip wall condition leads to a contact line singularity that is typically regularized with a Navier boundary condition, often requiring empirical fitting for the slip length. On the other hand, this paradox does not appear in phase-field models as the contact line moves through diffusive mass transport. In this work, we present a toy model that accounts for mass diffusion at the contact line within a sharp interface framework. This model is based on a theoretical relation derived from the Cahn-Hilliard equations, which links the total diffusive mass transport to the curvature at the wall. From an estimate of the chemical potential on a curved interface, we obtain an expression for the width of the highly curved region $\delta$ and the apparent angle. In the sharp interface model, we then introduce a fictitious boundary, displaced by a distance $\delta$ into the domain, where a Navier boundary condition is applied along a dynamic apparent contact angle that accounts for the local interface bending. The robustness of the model is assessed by comparing the toy model results with standard phase field ones on two cases: the steady state profiles of a liquid bridge between two plates moving in opposite directions and the transient behaviors of a drop spreading on a solid with a prescribed equilibrium angle. This offers a practical and efficient alternative to solve contact line problems at lower cost in a sharp interface framework with input parameters from phase field models.
\end{abstract}

\maketitle

\newpage
\section{Introduction}
\label{sec:Introduction}
The contact line formed at the intersection of a fluid-fluid and a solid boundary is a key element of many natural and several technological processes and its dynamics has been a subject of debate for several decades~\cite{Blake2002, Hocking_drop_spreading, Eggers_comment, Wilson2006, Yulli_review, Bonn2009,Snoeijer2013}. The classical no-slip boundary condition at a moving fluid interface leads to a force singularity at the contact line, as demonstrated Huh and Scriven~\cite{Huh1971}, implying that contact line motion would be impossible. 

In the context of continuum mechanics, in sharp interface models, one common approach is to introduce a small slip velocity at the contact line using the Navier boundary condition. The slip length, typically on the order of nanometers, is often associated with surface roughness or the mean free path of molecules~\cite{Lacis2020}. The Navier slip model combined with an equilibrium contact angle on a mesh-aligned boundary has been extensively studied across various physical setups using different numerical methods. For instance, a rapidly advancing contact line in a curtain coating scenario was simulated in a volume-of-fluid (VOF) framework~\cite{fullana_tomas_dynamic_2020}, different numerical methods with the same contact line model were tested in a capillary rise setup~\cite{grunding_comparative_2020}, a withdrawing plate was simulated using the VOF method with a mesh-dependent slip length~\cite{Afkhami2009}. Spreading droplets were explored with mesh-dependent slip~\cite{legendre_comparison_2015}, and the level-set method was employed for a spreading drop case, incorporating contact line hysteresis by varying the contact angle based on the contact line speed~\cite{spelt_level-set_2005}. Recently, a Generalized Navier Boundary Condition implemented within the same sharp interface framework as the present work was shown to regularize the logarithmic singularity at the contact line by allowing a deviation of the dynamic contact angle with respect to the equilibrium one through the uncompensated Young's stress~\cite{kulkarni_stream_2023, fullana2024consistenttreatmentdynamiccontact}. These models have in common that they require empirical information on the slip length, which can be viewed as a fitting parameter. 

On the other hand, in the phase-field method (PF), the Cahn-Hilliard equations are formulated from the thermodynamics of an immiscible two-component mixture and a phase function is used to represent the moving interface. The velocity field can satisfy a no-slip condition at the contact line, and the contact line moves due the diffusive mass transport that is present in a two-component system~\cite{jacqmin1999calculation, yue2010sharp}. While the phase-field method has the advantage of needing less empirical fitting, it is considerably more expensive. Although it has been shown that it is possible to match the results obtained on dynamic contact lines by fine-tuning the different parameters of the models~\cite{lacis_steady_2020,lacis2022}, an ideal model would mimic mass diffusion by a simple model; that is to obtain results essentially equivalent to the PF ones by doing much cheaper VOF simulations.  

We introduce a computational boundary at a specified short distance from the wall in the VOF simulation. There we apply a Navier condition for tangential velocity, and an apparent contact angle that accounts for the interface curvature at the wall. The basis for the model is a theoretical relation between the total diffusive mass transport across the interface and the interface curvature at the wall, that is derived from the Cahn-Hilliard equations. We compare full phase field simulations with VOF simulations using these boundary conditions. The input parameters for these VOF simulations are the same as for the full PF solutions. The VOF solutions reproduce the PF solutions, at a much lower cost, and within the framework of well established methods in CFD.

The paper is organized as follows. Section~\ref{sec:sharp} describes the sharp interface framework and Section~\ref{sec:diffuse} the phase field formulation. In Section~\ref{sec:toy} we derive the toy model for the dynamic contact line starting from the phase-field approximation of the incompressible two-phase Navier-Stokes equations. In Section~\ref{sec:results} we show a comparison of the toy model VOF results with standard PF ones on two different systems. Conclusion follows in Section~\ref{sec:conclusion}. 

\section{Sharp interface model}
\label{sec:sharp}
The volume-of-fluid method for representing fluid interfaces coupled with a flow solver is well-known to be suited for solving interfacial flows~\cite{Scardovelli1999,Popinet1999,tryggvason_scardovelli_zaleski_2011}. In the present work, we use the free software \href{http://basilisk.fr/}{Basilisk}, a platform for the solution of partial differential equations on adaptive Cartesian meshes~\cite{popinet_gerris_2003,popinet_accurate_2009,popinet_quadtree-adaptive_2015}.
We consider two incompressible, Newtonian and non-miscible fluids. We assume that there is no mass and heat transfer at the fluid-fluid interface, that the surface tension between the two fluids $\sigma$ is constant, and neglect gravity. The two fluids will be denoted $l$ and $g$ for liquid and gas phases, with densities and dynamical viscosities $\rho_l$ and $\rho_g$, $\mu_l$ and $\mu_g$ respectively.  
In the one-fluid formulation, the unsteady, incompressible, variable density Navier-Stokes equations read
\begin{align}
\dfrac{\partial \rho}{\partial t}+\mathbf{\nabla} \cdot(\rho \, \mathbf{u}) &= 0, \label{eq:NS1} \\
\rho\left(\dfrac{\partial \mathbf{u}}{\partial t}+\mathbf{u} \cdot \mathbf{\nabla} \mathbf{u}\right) &= -\mathbf{\nabla} P+\mathbf{\nabla} \cdot(2 \mu \mathbf{D})+ \mathbf{F}_\sigma, \label{eq:mom} \\
\mathbf{\nabla} \cdot \mathbf{u} &= 0, \label{eq:NS3}
\end{align}
with $\mathbf{u} \equiv \mathbf{u}(\mathbf{x}, t)$ the flow velocity, $P \equiv P(\mathbf{x}, t)$ the pressure, $\mathbf{D} = (\mathbf{\nabla} \mathbf{u} + \mathbf{\nabla} \mathbf{u}^T)/2$ the rate of deformation tensor, $\rho \equiv \rho(\mathbf{x}, t)$ the density, $\mu \equiv \mu(\mathbf{x}, t)$ the viscosity and $\mathbf{F}_\sigma$ the surface tension force. The surface tension term is modeled using the continuum surface force~\cite{brackbill_continuum_1992} so that $\mathbf{F}_\sigma = \sigma \, \kappa \, \nabla \chi $  with $\kappa$ the curvature, $\mathbf{n}$ the normal unit vector to the interface and $\chi$ the Heaviside function 
\begin{equation} \label{eq:chi}
\chi(\mathbf{x}, t) = \left\{
\begin{array}{rl} 
0, \quad & \text{in the gas phase}, \\ 
1, \quad & \text{in the liquid phase},
\end{array}\right.
\end{equation}
also used to locate the interface and define the viscosity and density~\cite{tryggvason_scardovelli_zaleski_2011}.
The volume fraction $c(\mathbf{x}, t)$ corresponding to the average value of $\chi$ in each cell is used to replace the advection of the density  \eqref{eq:NS1} by the following advection equation:
\begin{equation} \label{eq:advVOF}
\dfrac{\partial c}{\partial t} + \mathbf{u} \cdot \mathbf{\nabla} c = 0.
\end{equation}
The local density and viscosity are then defined from the local volume fraction $c$
\begin{equation}\label{eq:rho7}
\begin{aligned}
\rho(c) \equiv c \, \rho_l + (1 - c) \, \rho_g, \\
\mu(c) \equiv c \, \mu_l + (1 - c) \, \mu_g. 
\end{aligned}
\end{equation}
The incompressible Navier-Stokes equations are approximated using a time-staggered approximate projection method on a Cartesian grid. The advection term is discretized with the explicit and conservative Bell–Colella–Glaz second-order unsplit upwind scheme~\cite{bell_second-order_1989}. For the discretization of the viscous diffusion term, a second-order Crank–Nicholson fully-implicit scheme is used. Spatial discretization is achieved using a quadtree (in two dimensions) adaptive mesh refinement on collocated grids. The adaptive mesh technique is based on the wavelet decomposition of the variables such as the velocity field or the volume fraction and is efficient and critical to the success of the simulations as it allows high resolution only where needed, therefore reducing the cost of calculation.

All the variables are co-located at the center of each square discretization volume. Consistently with a finite-volume formulation, the variables are interpreted as the volume-averaged values for the corresponding discretization volume. To solve the advection equation the geometrical VOF scheme is used and proceeds in two steps: (i) Interface reconstruction (ii) Geometrical of flux estimation and interface advection. The reconstruction is a ‘piecewise linear interface calculation’ (PLIC), followed by a Lagrangian advection.

The resolution of the surface tension term is directly dependent on the accuracy of the curvature calculation. The Height-Function methodology is a VOF-based technique for calculating interface normals and curvatures~\cite{afkhami_height_2008, afkhami_height_2009}. About each interface cell, fluid ‘heights’ are calculated by summing fluid volume in the grid direction closest to the normal of the interface. In two dimensions, a $3 \times 7$ stencil around an interface cell is constructed and the heights are evaluated by summing volume fractions vertically 
\begin{equation}
h_{j}=\sum_{k=i-3}^{k=i+3} c_{j, k} \Delta_{\text{V}},
\end{equation}
with $c_{j, k}$ the volume fraction and $\Delta_{\text{V}}$ the grid spacing. The heights are then used to compute the the interface normal $\mathbf{n}$ and the curvature $\kappa$
\begin{equation}\label{eq:basilisk_heights}
\begin{array}{c}{\mathbf{n}=\left(h_{x},-1\right)}, \\  \\ {\kappa=\dfrac{h_{x x}}{\left(1+h_{x}^{2}\right)^{3 / 2}}}, \end{array}
\end{equation}
where $h_{x}$ and $h_{xx}$ are discretized using second-order central differences.
This method is also used to model the contact angle at the triple point. The orientation of the interface, characterized by the contact angle -- the angle between the normal to the interface at the contact line and the normal to the solid boundary -- is imposed in the contact line cell. It is important to note that a numerical specification of the contact angle affects the overall flow calculation in two ways: (i) it defines the orientation of the VOF reconstruction in cells that contain the contact line; (ii) it influences the calculation of the surface tension term by affecting the curvature computed in cells at and near the contact line. 
The boundary conditions in the VOF toy model, derived in Section~\ref{sec:toy}, will result in a simple slip model coupled to a dynamic contact angle. The slip model, or Navier boundary condition, has been implemented and used in various studies within this numerical framework~\cite{fullana_tomas_dynamic_2020,lacis_steady_2020}. In two dimensions, the Navier boundary condition reads
\begin{align}\label{eq:SLIP}
\begin{split}
u_x|_{y=0} + \left.\lambda \dfrac{\partial u_x}{\partial y}\right|_{y=0} & = U, \\
u_y|_{y=0}  &= 0,
\end{split}
\end{align}{}
with $u_x|_{y=0}$ and $u_y|_{y=0}$ the $x$ and $y$ component of the velocity at the solid boundary, $\lambda$ the slip length, and $U$  the prescribed velocity of the moving solid. Note that \eqref{eq:SLIP} is written with the boundary located at $y=0$; the location of the computational boundary will be lifted in the toy model (Section~\ref{sec:toy}) such that the boundary is located at $y=+\lambda$. Moreover, it is important to remark that the accuracy of the numerical method is dependent on the resolution of the smallest length scale, the slip length $\lambda$. In the following numerical setups, the grid size will be chosen such that $\lambda / \Delta_{\text{V}} \simeq 4$. \change{This particular choice is based on our experience in contact line systems with the VOF method where sufficient resolution (i.e. grid size smaller than the slip length) is necessary to achieve grid-converged results~\cite{fullana_tomas_dynamic_2020}.}

\section{Diffuse interface model}
\label{sec:diffuse}

In the phase-field method, the governing equations are derived from the thermodynamic potentials of the system, together with the assumption of a surface energy associated with an interface~\cite{cahn1958free, jacqmin2000contact, yue2010sharp}. It is, therefore, possible to consider different physical situations with relative ease. It is also straightforward to implement numerically, since interfaces are not tracked explicitly. Instead, a variable is introduced that has different constant values in the two phases with a steep transition between the two in the diffuse interface~\cite{jacqmin2000contact,carlson2012thesis}. One of the major drawbacks of this method, however, is that the width of the interface $\epsilon$ must be small to match the proper interface dynamics.

Similarly to the one-fluid formulation, the phase-field model introduces a phase variable $C(\mathbf{x},t)$ ranging from $1$ to $-1$. This phase-field variable $C(\mathbf{x},t)$ is governed by a convection-diffusion equation
\begin{equation}\label{eq:PF0}
\dfrac{\partial C}{\partial t} = \nabla \cdot (-\mathbf{F}_d - \mathbf{F}_c),
\end{equation}
where $\mathbf{F}_d$ is the diffusive flux and $\mathbf{F}_c$ is the convective flux. The latter, in the context of incompressible flows, takes the simple form 
\begin{equation}\label{eq:PF1}
\mathbf{F}_c = \mathbf{u} \cdot \nabla C,
\end{equation}
and the diffusive flux is
\begin{equation}\label{eq:PF2}
\mathbf{F}_d = - M \: \nabla \phi,
\end{equation}
where $M$ is a proportionality coefficient called the phase-field mobility and $\phi$ the chemical potential defined as
\begin{equation}\label{eq:chempot}
\phi = \beta \Psi'\left(C \right) - \alpha \nabla^2 C.
\end{equation}
In the chemical potential, we have two parameters $\alpha$ and $\beta$, which are related to the surface tension $\sigma$ and the characteristic thickness of the diffuse interface $\epsilon$ as
\begin{equation}\label{eq:PF3}
\sigma = \dfrac{2}{3} \sqrt{2 \alpha \beta}, \quad \quad \epsilon = \sqrt{\dfrac{\alpha}{\beta}}.
\end{equation}
In addition, it contains the derivative of the standard double-well potential 
\begin{equation}\label{eq:doublewell}
\Psi \left(C\right) = \dfrac{\left( C+1 \right)^2\left( C-1 \right)^2}{4}.
\end{equation}
With the definitions above \eqref{eq:PF0}-~\eqref{eq:doublewell}, the convection-diffusion equation, referred to as the Cahn-Hilliard (\citeyear{CH1958}) equation, can be written as
\begin{equation}\label{eq:cahn-hil-dim}
\displaystyle \dfrac{\partial C}{\partial t} = \nabla \cdot \left[ M \nabla \left(\beta \Psi'\left(C \right) - \alpha \nabla^2 C \right) \right] - \mathbf{u} \cdot \nabla C.
\end{equation}
Equation~\eqref{eq:cahn-hil-dim} is the phase-field analog of~\eqref{eq:advVOF}. The boundary conditions for this fourth-order partial differential equation will be detailed in our study of dynamic wetting by the phase-field method. Taking the same notation as in~\eqref{eq:mom}, the momentum equation in the Navier-Stokes equations becomes
\begin{equation}\label{eq:momPF}
\rho\left(\dfrac{\partial \mathbf{u}}{\partial t}+\mathbf{u} \cdot \nabla \mathbf{u}\right)=-\nabla P+\nabla \cdot(2 \mu \mathbf{D}) + \phi \nabla C,
\end{equation}
where $ \phi \nabla C$ corresponds to the surface tension force and acts over the diffuse interface region. 
Moreover, the density and viscosity are now defined through the phase-field variable $C\ $ as
\begin{equation}\label{eq:rho7PF}
\begin{aligned} 
\displaystyle \rho({C}) & \equiv \rho_{l} \dfrac{C + 1}{2}+\rho_{g} \dfrac{C - 1}{2}, \\ 
\displaystyle \mu({C}) & \equiv \mu_{l} \dfrac{C + 1}{2}+\mu_{g} \dfrac{C - 1}{2}.
\end{aligned}
\end{equation} 
We now state the boundary conditions in the presence of a contact line on a solid, characterized by its equilibrium angle (the angle between the fluid-fluid interface and the wall). The convection-diffusion equation~\eqref{eq:cahn-hil-dim} is a fourth-order partial differential equation and requires two boundary conditions. First, we impose a non-equilibrium wetting condition~\cite{jacqmin2000contact,qian2003molecular} on the solid wall,
\begin{equation}
- \mu_f \, \epsilon \left( \dfrac{\partial C}{\partial t} + \mathbf{u} \cdot \nabla C \right) = \alpha \nabla C \cdot \mathbf{n} - \sigma \cos \theta_\mathrm{e} \: g'\left( C \right),
\end{equation}
where $\mu_f$ is a contact line friction parameter, having the same units as bulk dynamic viscosity. Here, $\theta_\mathrm{e}$ is the equilibrium contact angle and 
\begin{equation}
g\left(C\right) = \dfrac{1}{4} (2 - 3 C + C^3),
\end{equation}
is a switching function describing a smooth transition between both phases. The unit normal vector $\mathbf{n}$ is directed from the fluid to the surrounding solid.

If one sets $\mu_f = 0$, the contact angle is always enforced to the equilibrium angle $\theta_\mathrm{e}$. This will be the case in the toy model when comparing PF to VOF simulations. Non-zero contact line friction allows the dynamic contact angle to evolve naturally as a function of contact line speed. The second boundary condition for the phase function is a zero diffusive flux of chemical potential through the boundaries 
\begin{equation}
\nabla \phi \cdot \mathbf{n} = 0.
\end{equation}
The phase-field equations are written in a weak form, and solved using the open-source finite-element software \emph{FreeFEM} by Hecht, Pironneau and others~\cite{FREEFEM,freefem-}, allowing the specification of a finite-element weak form as input. Linear elements were used for the PF variables, while the fluid flow was resolved using Taylor–Hood elements (quadratic for velocity and linear for pressure). The numerical code is available in the Github repository \href{https://github.com/UgisL/FreeFEM-NS-CH}{FreeFEM-NS-CH}. Similarly to the VOF implementation, the accuracy of the method depends on the resolution of the smallest length scale of the system, the interface width $\epsilon$. In the PF setups, the grid size will be chosen such that $\epsilon / \Delta_{\text{P}} \simeq 16$.

\section{The toy model: relating mass flux to curvature}
\label{sec:toy}
We start by formulating the dimensionless incompressible two-phase Navier-Stokes equations in the phase-field approximation. As reference quantities we choose $U$, $L$, $L/U$ and $\mu_l U /L$ for velocity, length, time and pressure, with $L$ denoting a macroscopic length, for example the droplet radius, and $U$ the contact line speed. The dimensionless numbers that appear are: the Reynolds number $\Re = \rho_l U L/\mu_l$, the Peclet number $\Pe= U L / D$, where $D=M \sigma / \epsilon$ is the mass diffusion coefficient, and the capillary number $\Ca= \mu_l U / \sigma$. Lastly, the Cahn number -- or dimensionless interface width -- is defined as $\Cn = \epsilon/L$ and $\Ca_F= \mu_F U / \sigma$ is the line friction capillary number, based on the line friction parameter $\mu_F$. In the rest of this study, we consider the line friction coefficient to be zero ($\mu_F = 0$), in order to enforce the dynamic contact angle to the equilibrium one. The dimensionless equations therefore read
\begin{align}\label{eq:NSPF_nondim}
\rho_l  \left( \dfrac{\partial \mathbf{u}}{\partial t} + \mathbf{u} \cdot \nabla \mathbf{u} \right) = -\dfrac{1}{\Re} \nabla P^* + \dfrac{1}{\Re} \nabla^2 \mathbf{u} - \dfrac{1}{\Re \: \Cn \: \Ca} C  \nabla \phi,
\end{align}
\begin{equation} \label{eq:diffueq}
\dfrac{\partial C}{\partial t}+ \mathbf{u} \cdot \nabla C  = \dfrac{1}{\Pe} \nabla^2 \phi,
\end{equation}
with
\begin{equation}\label{eq:chempotdef}
\phi=-\Cn^2 \nabla^2 C  +  \Psi'\left(C \right),
\end{equation}
and
\begin{equation}\label{eq:chempotdefprime}
\Psi' \left (C \right)= (C^2-1) C.
\end{equation}
Here the pressure has been replaced by the reduced pressure $P^* = P - C \phi  /(\Cn \Ca)$, which is continuous across the interface at equilibrium.

\begin{figure}
    \centering
    \includegraphics[width=0.7\linewidth]{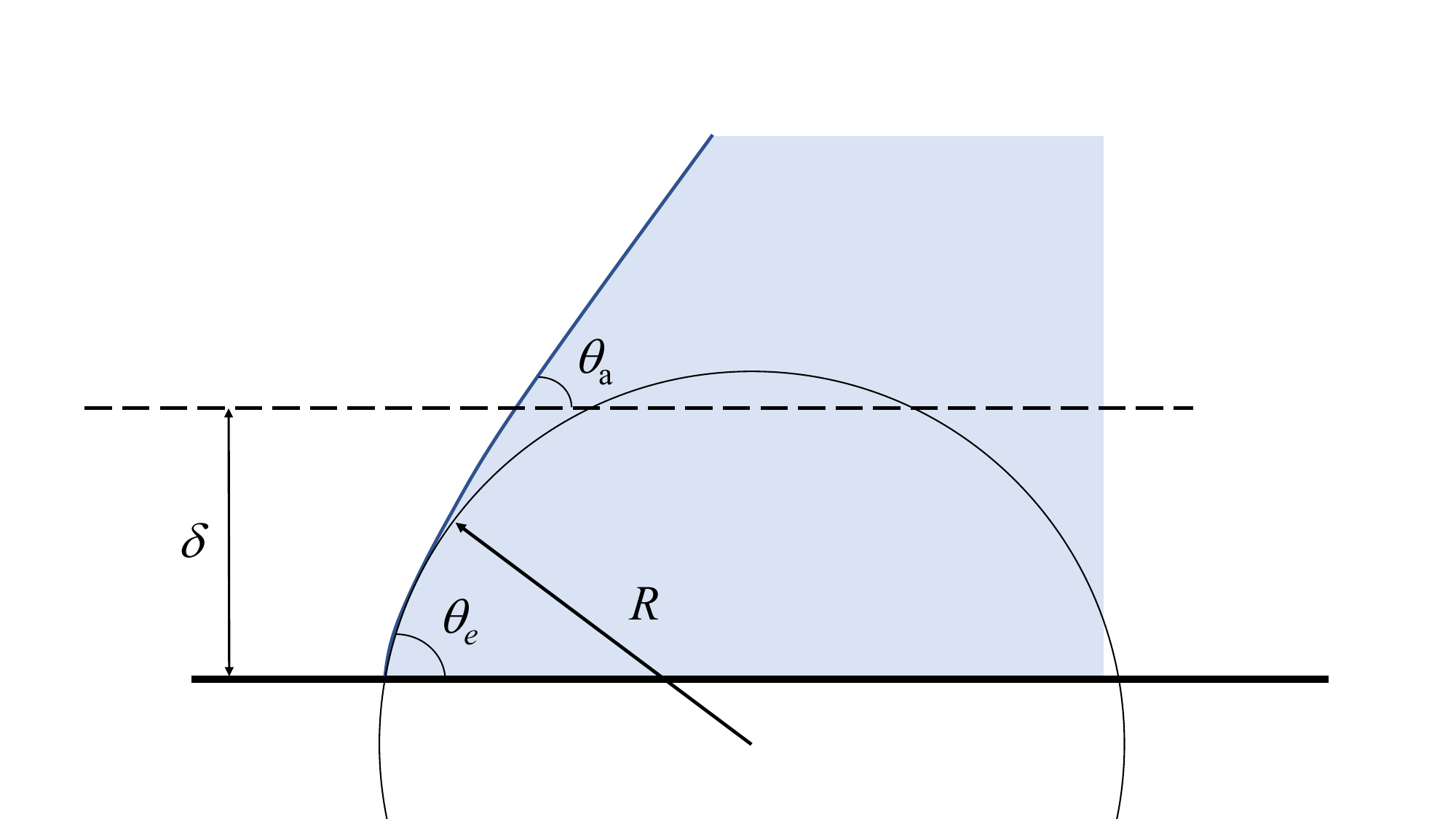}
    \caption{Sketch of the region of high curvature close to a moving contact line.}
    \label{fig:curvaturesketch}
\end{figure}

The wetting boundary condition, with $\mu_F = 0$, is now
\begin{equation}\label{eq:wettingBC}
\dfrac{3}{2\sqrt{2}} \Cn \nabla C \cdot \mathbf{n} = \cos (\theta_\mathrm{e}) g'(C).
\end{equation}
In what follows, we assume that there is a small region near the contact line where the interface has locally a high curvature, as shown in Fig.~\ref{fig:curvaturesketch}. We assume the contact angle at the wall to be kept at the nominal equilibrium value $\theta_\mathrm{e}$, but that it will adjust to an apparent contact angle $\theta_\mathrm{a}$ over a short distance $\delta$. We will now proceed to derive estimates for the chemical potential on a curved interface, and from this, the local curvature at the moving contact line. Using this, we will arrive at semi-quantitative expressions for the width of the highly curved region, and the value of the apparent contact angle $\theta_\mathrm{a}$.
Assume a length scale $\deltayzf$, a viscous balance in the momentum equation, and that the velocity is properly scaled (dimensionless $|\mathbf{u}| \sim 1)$. Estimates of the relevant terms in~\eqref{eq:NSPF_nondim}--\eqref{eq:diffueq} then give

\begin{equation}\label{eq:delta_toyNS}
\dfrac{1}{\Re} \: \dfrac{1}{\deltayzf^2} \quad \sim  \quad  \dfrac{1}{\Ca \Cn \Re} \: \dfrac{\phi}{\deltayzf},
\end{equation}
and
\begin{equation}\label{eq:delta_toyDiff}
 \dfrac{1}{\deltayzf} \quad \sim \quad \frac{1}{\Pe} \: \frac{\phi}{\deltayzf^2}.
\end{equation}
Solving for $\deltayzf$ yields 
\begin{equation}\label{eq:delta_toy}
\deltayzf \quad \sim \quad  \sqrt{ \dfrac{\Ca \Cn}{\Pe}} = \dfrac{\sqrt{\mu_l M}}{L},
\end{equation}
which agrees with the scaling mentioned by several authors previously~\cite{jacqmin2000contact,QIAN:2006hl, yue2010sharp}. 
 Briant et al.~\cite{briantyeomans2004} have given a somewhat different expression  $(\mu_l M \epsilon^2)^{1/4}  /  L$. This however involves the interface thickness in the estimates of the concentration gradients, while the length scale in \eqref{eq:delta_toy} characterizes the length scale for diffusion in the contact line region, away from the interface.

We now derive an expression for the chemical potential on a locally curved interface. Assuming a planar geometry and a circular interface of radius $R$, the chemical potential as defined in~\eqref{eq:chempotdef} can be written as
\begin{equation}\label{eq:pot_toy}
\phi=-\Cn^2 \left(\dfrac{\partial^2 C}{\partial r^2} + \dfrac{1}{r} \dfrac{\partial C}{\partial r} \right)  +  \Psi' \left(C \right).
\end{equation}
$C$ is expected to approach $+1$ inside the droplet, and $-1$ outside. Since the dimensionless interface width is $\Cn \ll 1$, the solution to this equation has a boundary layer character, and we introduce a new radial coordinate $s$ such that $r=R+\Cn \cdot s$ and expand~\eqref{eq:pot_toy} at zeroth and first order 
\begin{equation}\label{eq:zerothorder0}
C = C_0 + \Cn \cdot C_1 + O(\Cn^2),
\end{equation}
\begin{equation}\label{eq:firstorder0}
\phi=\phi_0 + \Cn \cdot \phi_1 + O(\Cn^2).
\end{equation}
The problems for the zeroth and first order are now
\begin{equation}\label{eq:zerothorder}
\phi_0= - \dfrac{\partial^2 C_0}{\partial s^2}  +  \Psi'\left(C_0 \right),  
\end{equation}
and
\begin{equation}\label{eq:firstorder}
\phi_1= - \left( \dfrac{\partial^2 C_1}{\partial s^2} - \dfrac{1}{R} \dfrac{\partial C_0}{\partial s} \right)  +  C_1  \Psi'' \left(C_0 \right).
\end{equation}
The zeroth order~\eqref{eq:zerothorder} gives the equilibrium solution as 
\begin{equation}\label{eq:zerothsoln}
C_0= -\tanh(s/ \sqrt{2}), \text{ with   } \phi_0 = 0.
\end{equation}
Multiplying~\eqref{eq:firstorder} by $\partial C_0 / \partial s$ and partially integrating, a solvability condition is obtained for the first-order value of the chemical potential
\begin{equation}\label{eq:solvability}
\displaystyle \int \phi_1 C_{0,s} ds=-\int (C_{0,s} C_{1,ss} + C_1 (\Psi'(C_0))_{,s}) ds -   \dfrac{1}{R} \int (C_{0,s})^2  ds,  
\end{equation}
where the subscript denotes derivative with respect to $s$. The chemical potential is continuous at the interface, so the first order chemical potential $\phi_1$ is a constant in~\eqref{eq:firstorder} and~\eqref{eq:solvability}. By partially integrating the first term on the right hand side that term is seen to be zero. Using that  
\begin{equation}\label{eq:chemconditions}
\int C_{0,s} ds = -2 \quad \text{and} \quad \int (C_{0,s})^2  ds = \dfrac{2 \sqrt{2}}{3},  
\end{equation} 
we obtain the value of the chemical potential at the interface $\phi_i$
\begin{equation}\label{eq:chempot_interface}
\phi_i = \Cn \phi_1 = \dfrac{1}{2} \dfrac{2 \sqrt{2}}{3} \dfrac{\Cn}{R}.
\end{equation} 
Turning now to~\eqref{eq:diffueq}, which expresses the mass balance of the concentration $C$, the solution to this equation varies on two different length scales: one is the interface width and the other a larger length scale characterizing the region around the contact line where mass diffusion restores the concentration to its equilibrium value, i.e. the Yue-Zhou-Feng length-scale $\deltayzf$ in~\eqref{eq:delta_toy}.
In order to establish a quantitative model, we now rewrite the scaling of~\eqref{eq:diffueq} in~\eqref{eq:delta_toyDiff} as an equality
\begin{equation}\label{eq:delta_toyDiff_eq}
 b \: \dfrac{2}{\delta} \quad = \quad \frac{1}{\Pe} \: \frac{\phi_i}{\delta^2}.
\end{equation}
Here we introduced a non-dimensional factor $b$ of order unity, which accounts for the  shape of the velocity profile at the contact line. We note that the full equation~\eqref{eq:diffueq}, when re-scaled according to~\eqref{eq:delta_toy}, will be a non-dimensional equation, independent of model parameters. As $b$ is determined from a solution of this re-scaled system, we conclude that $b$ is independent of the model parameters; with the exception of the equilibrium contact angle $\theta_\mathrm{e}$. For partially wetting systems where $\theta_\mathrm{e}$ is not close to 0 or $180\degree$ we expect this dependence to be weak.

We now consider the Navier-Stokes equations~\eqref{eq:NSPF_nondim}. Assuming a balance between the viscous term and the capillary term, we estimate the relevant terms in the Navier-Stokes equation as 
\begin{equation}\label{eq:toyNS}
\dfrac{1}{\Re} \, \dfrac{\tau}{\delta} \quad = \quad 
\dfrac{1}{\Ca \Cn \Re}\,
\phi_i \, \dfrac{2}{\delta}.
\end{equation}
Here a dimensionless parameter $a$ of order unity is introduced to account for the viscous stress 
$\tau = a \, u/\delta$. As discussed above in relation to the non-dimensional parameter $b$, $a$ is similarly determined from a re-scaled equation, free from model parameters, as long as the flow is considered inertialess. However, if the two fluid regions have different viscosities, the viscosity ratio does enter. From our numerical experiments however this dependence appears weak.   

Solving now for $R$ and $\delta$ from~\eqref{eq:chempot_interface}--\eqref{eq:toyNS} yields the local curvature as a function of the flow parameters 
\begin{equation}\label{eq:Rab}
R = \dfrac{1}{2} \dfrac{2 \sqrt{2}}{3} \dfrac{1}{\sqrt{a \, b}} \sqrt{\dfrac{\Cn}{\Pe \Ca}}.
\end{equation}
\begin{equation}\label{eq:delta_ab}
\delta = \dfrac{1}{2}   \sqrt{\dfrac{a}{b}} \sqrt{\dfrac{\Ca\Cn }{\Pe}}.
\end{equation}
 
Throughout the treatment above, the viscosity is taken as the higher liquid viscosity, in the case of unequal viscosities in the two regions, and as noted above, the dependence of our results on the viscosity ratio is weak. This may seem to be in contrast with the observation of Yue et al.~\cite{yue2010sharp} (Figs. 9 and 11), who note that the distance from the wall to the stagnation point near the contact line (the boundary layer thickness) should be based on the geometric mean viscosity $\mu_e= \sqrt{\mu_l \,  \mu_a}$. Here, however, the interpretation of the boundary layer thickness is not the distance to the stagnation point per se, but rather estimates the width of the highly curved region. One explanation could be that, as is clear from flow fields in the phase-field simulations, the stagnation point is displaced into the low viscosity region when the viscosities differ. For the analysis in the present work, it is the action of the viscous stresses and the shape of the velocity profile in the contact line region that is important, and not the location of the stagnation point as such.


The local radius of curvature $R$ at the contact line allows us to estimate how the local angle $\theta_\mathrm{a}$ varies as function of the distance to the wall. Assuming a locally circular interface of radius $R$ according to~\eqref{eq:Rab}, some geometrical considerations show that the angle $\theta_\mathrm{a}$, taken at a distance $\delta$ from the wall according to~\eqref{eq:delta_ab}, is given by the following relation: 
\begin{equation}\label{eq:toyangle}
\cos(\theta_\mathrm{a}) = \cos(\theta_\mathrm{e}) + \dfrac{\delta}{R}  = \cos(\theta_\mathrm{e}) + \dfrac{3}{2 \sqrt{2}} \Ca a.
\end{equation} 
In what follows we will use~\eqref{eq:toyangle} to calculate the apparent contact angle $\theta_\mathrm{a}$ from the local capillary number $\Ca$, and apply this at a fictitious boundary displaced by the distance $\delta$ into the flow domain, as indicated in Fig. \ref{fig:curvaturesketch}.

\section{Results}
\label{sec:results}
\subsection{Sheared drop}
\label{sec:sheared_drop}

\begin{figure}
    \centering
    \subfloat[]{\includegraphics[width=0.6\linewidth]{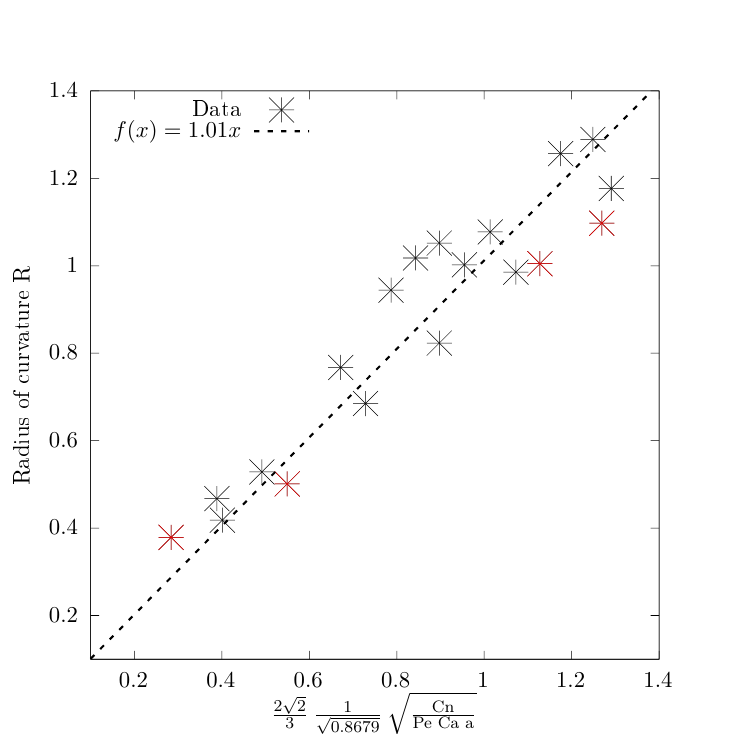}}
    \subfloat[]{
    \centering
    \begin{minipage}[b][0.6\linewidth][t]{.4\linewidth}
    \vspace{0.40cm}
\centering
Pe = 300  Cn = 0.02
\includegraphics[width=1\linewidth]{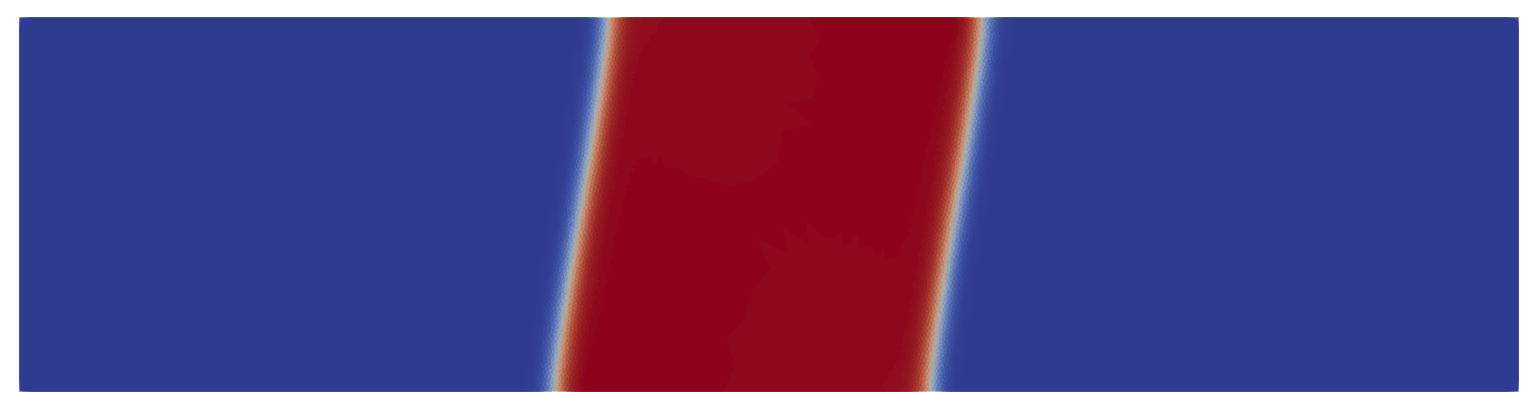}

Pe = 160  Cn = 0.04
\includegraphics[width=1\linewidth]{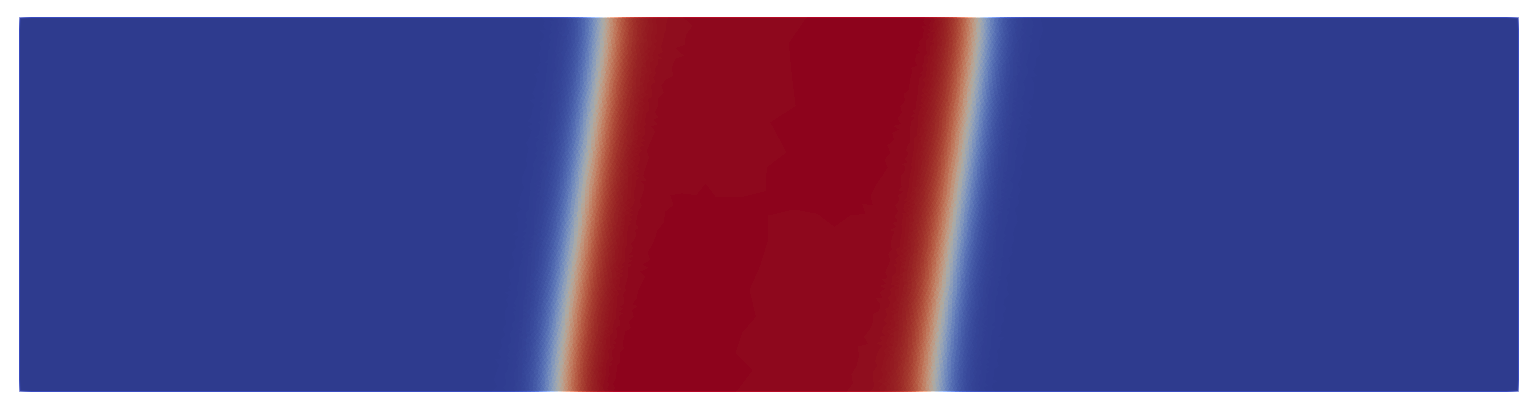}

Pe = 30  Cn = 0.02
\includegraphics[width=1\linewidth]{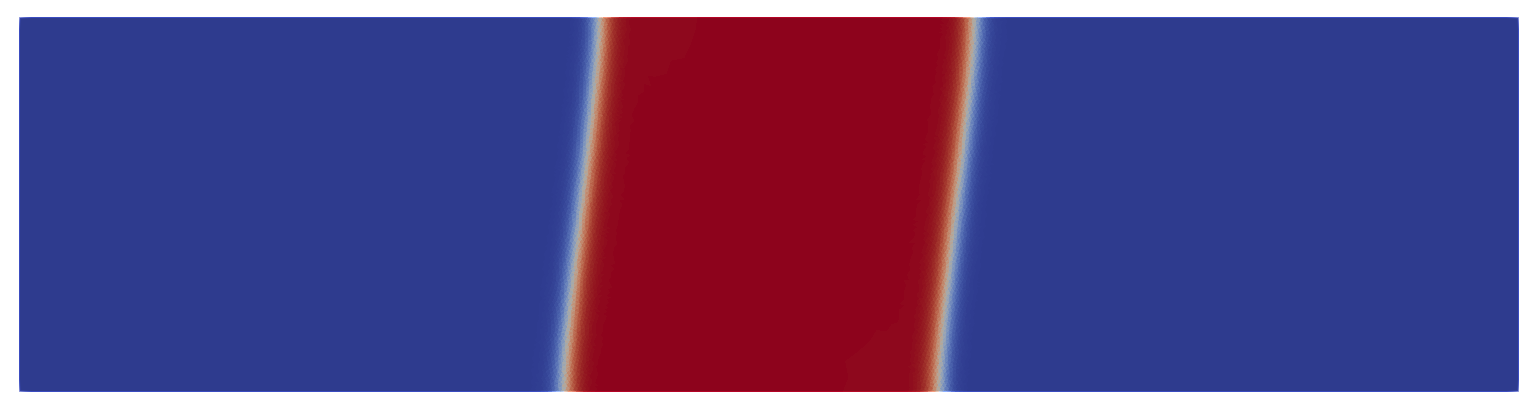}

Pe = 26  Cn = 0.04
\includegraphics[width=1\linewidth]{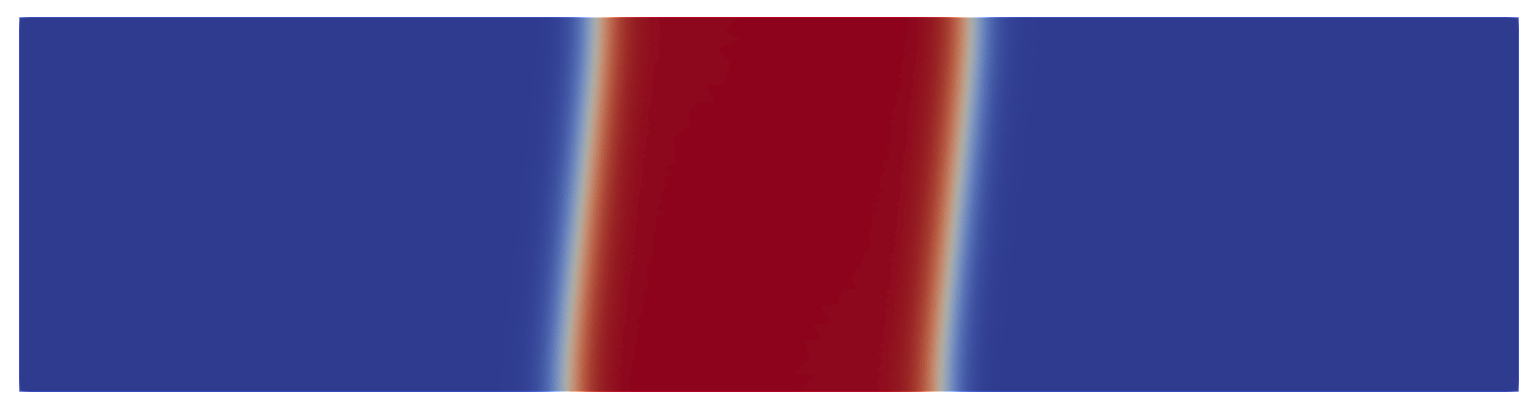}
\end{minipage}}%
    \caption{Correlation between the radius of curvature $R$ and~\eqref{eq:Rab}. (a) Each simulation corresponds to a unique set of Peclet ($\Pe$) and Cahn ($\Cn$) numbers while maintaining a constant capillary number ($\Ca$). \change{The toy model parameters are set to $a = 2.5$ and $b = 0.2170$}, and \change{the} fitted line demonstrates a clear one-to-one relationship between these two \change{quantities}. (b) Steady-state solutions of phase-field (PF) simulations for cases highlighted in red in the left panel.}
    \label{fig:a_factor}
\end{figure}

In the first test case, we study the steady shape of a nanoscopic droplet confined between two moving walls similarly to Lacis et al.~\cite{lacis_steady_2020, lacis2022}. Initially, we use the phase-field model to solve the system in order to validate the correlation between the radius of curvature and the flow parameters, as expressed in~\eqref{eq:Rab}. We choose a constant capillary number $\Ca = 0.0212$, which is below the critical capillary number, while systematically varying the Peclet and Cahn numbers. \change{The equilibrium angle is set to $\theta_e = 90\degree$ for both the advancing and receding contact lines, and} the geometrical factor $a = 2.5$ is kept fixed. 
For each simulation run, we compute the radius of curvature $R$ once the system reaches steady state, using the mean curvature at the contact line. The results, illustrated in Fig.~\ref{fig:a_factor}, demonstrate a remarkable agreement between $R$ and~\eqref{eq:Rab}, establishing a clear one-to-one relationship between these two \change{quantities}. \change{In this preliminary analysis, the parameter $b = 0.2170$ is fixed (see Appendix~\ref{appendixB} for details).} In Section~\ref{sec:spreading_drop}, we will further substantiate the appropriateness of this choice in the context of a spreading droplet.

\begin{figure}
\centering
\subfloat[]{
\centering
\includegraphics[width=0.15\linewidth]{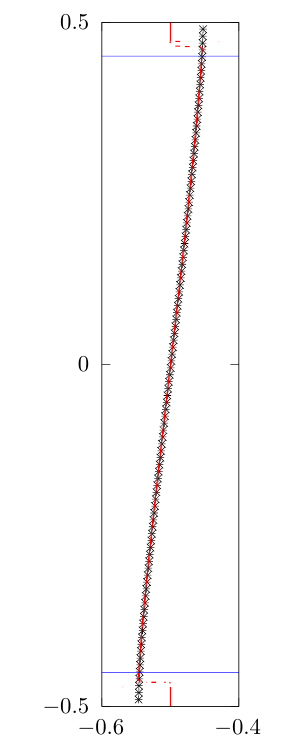}
}
\subfloat[]{
\centering
\includegraphics[width=0.15\linewidth]{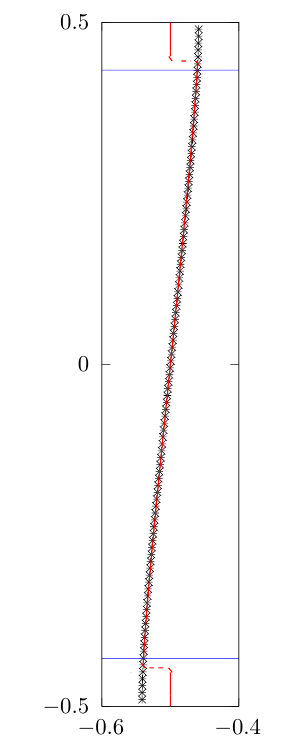}
}
\subfloat[]{
\centering
\includegraphics[width=0.15\linewidth]{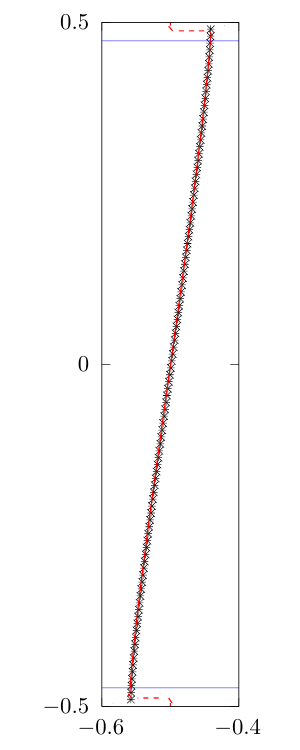}
}
\subfloat[]{
\centering
\includegraphics[width=0.15\linewidth]{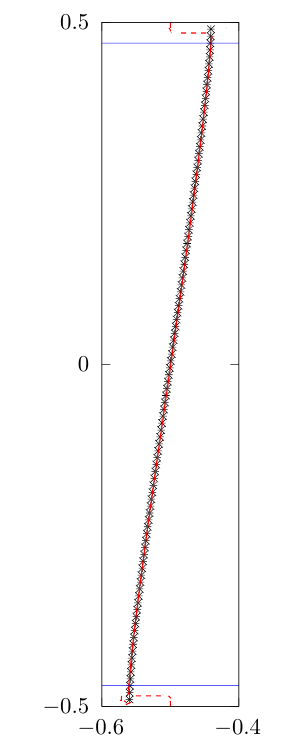}
}
\subfloat[]{
\centering
\includegraphics[width=0.15\linewidth]{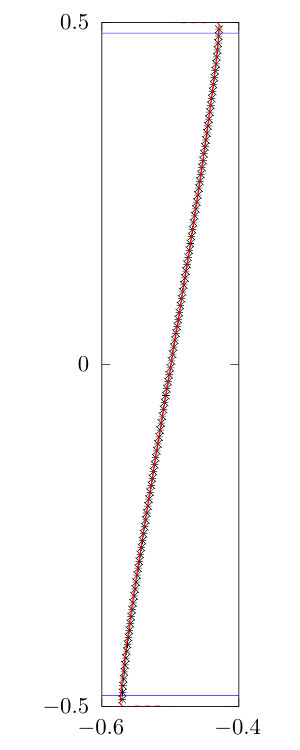}
}
\subfloat[]{
\centering
\includegraphics[width=0.15\linewidth]{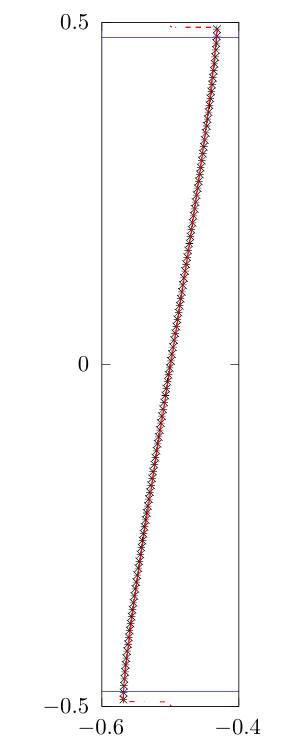}
}
\caption{Comparison between the phase-field left interface (black points) and the volume-of-fluid interface (red line) for six distinct sets of Peclet ($\Pe$) and Cahn ($\Cn$) numbers. Panels (a) to (f) correspond to the following parameter sets: ($\Pe = 30, \: \Cn = 0.02$), ($\Pe = 30, \: \Cn = 0.04$), ($\Pe = 100, \: \Cn = 0.02$), ($\Pe = 160, \: \Cn = 0.04$), ($\Pe = 300, \: \Cn = 0.02$), and ($\Pe = 300, \: \Cn = 0.04$).}
    \label{fig:PF_VOF comparison_prof}
\end{figure}

\begin{figure}
\centering
\subfloat[$\Pe = 30$ $\Cn = 0.02$]{
\centering
\includegraphics[width=0.3\linewidth]{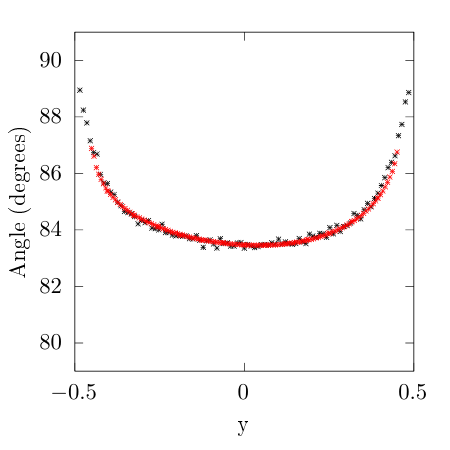}
}
\subfloat[$\Pe = 30$ $ \Cn = 0.04$]{
\centering
\includegraphics[width=0.3\linewidth]{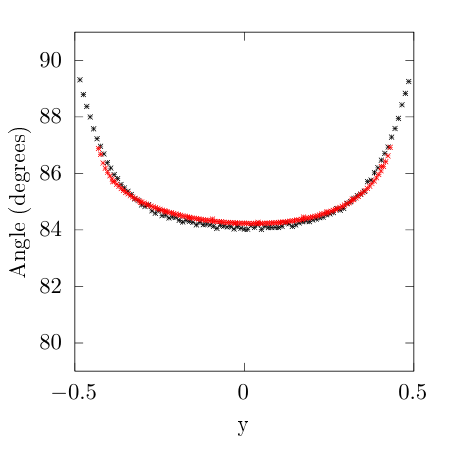}
}
\subfloat[$\Pe = 100$ $ \Cn = 0.02$]{
\centering
\includegraphics[width=0.3\linewidth]{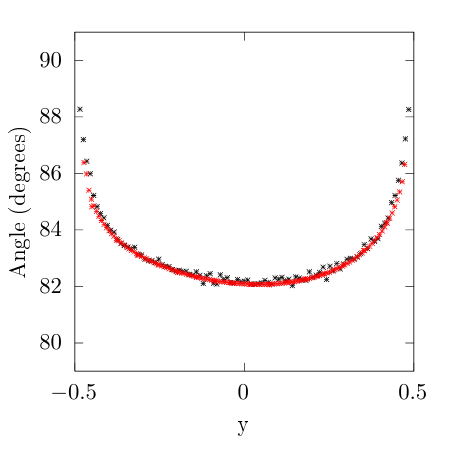}
}

\subfloat[$\Pe = 160$ $ \Cn = 0.04$]{
\centering
\includegraphics[width=0.3\linewidth]{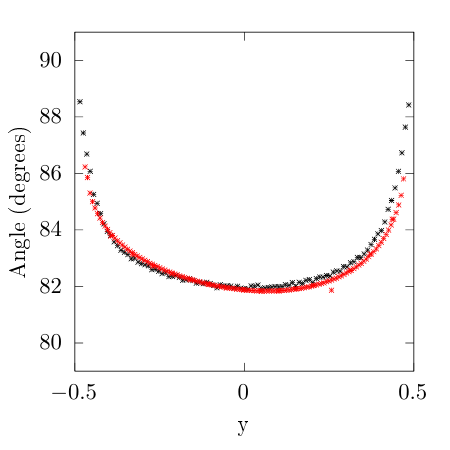}
}
\subfloat[$\Pe = 300$ $ \Cn = 0.02$]{
\centering
\includegraphics[width=0.3\linewidth]{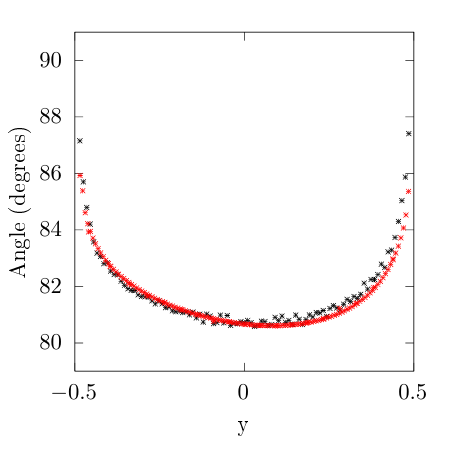}
}
\subfloat[$\Pe = 300$ $ \Cn = 0.04$]{
\centering
\includegraphics[width=0.3\linewidth]{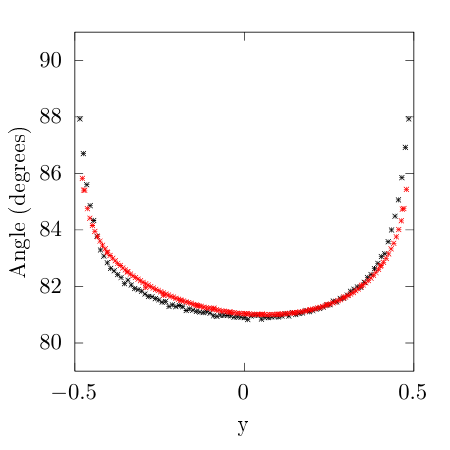}
}
\caption{Comparison between the phase-field angle along the left interface (black points) and the volume-of-fluid angle (red \change{symbols}) for six distinct sets of Peclet ($\Pe$) and Cahn ($\Cn$) numbers.}
    \label{fig:PF_VOF comparison_angle}
\end{figure}

We now test our toy model on this setup. Using the relations described in Section~\ref{sec:toy}, we compare full PF simulations -- in the sense that the full domain is solved -- with VOF simulations using the toy model approximation. In the toy model, the bottom boundary is now a computational boundary that is located a distance $+ \delta$ away from the wall. The domain comprised between the wall $y = 0$ and the computational boundary $y = \delta$ is not simulated. There, we apply the dynamic angle relation~\eqref{eq:toyangle}, where the apparent angle $\theta_\mathrm{a}$ is now a dynamic contact angle $\theta_d$ 
\begin{equation}\label{eq:toyangle2}
\theta_d = \cos^{\shortminus 1} \left(\cos \: \theta_\mathrm{e} + \dfrac{3}{2 \sqrt{2}} \Ca a \right),
\end{equation}
with $\Ca$ the contact line capillary number. Moreover, in order to simulate a true no-slip condition at the wall, we use $\delta$ as the slip length for the tangential velocity (again located at the computational boundary)
\begin{equation}\label{eq:toyslip}
u_x|_{y = \delta} - \delta \dfrac{\partial u_x }{\partial y}|_{y = \delta} = \pm \, U_w,
\end{equation}
with $U_w$ the wall velocity. Fig.~\ref{fig:PF_VOF comparison_prof} shows the resulting steady state interfaces for the left side of a sheared droplet, for a range of $\Pe$ and $\Cn$. The full PF simulations are compared to the corresponding VOF results, using equations~\eqref{eq:toyangle2} and~\eqref{eq:toyslip} as boundary conditions, with good agreement. A more discerning test is shown in Fig.~\ref{fig:PF_VOF comparison_angle} where the local angle that the interface makes with the horizontal is shown as a function of vertical position, for the same six cases as in Fig.~\ref{fig:PF_VOF comparison_prof}. The details of the profile shapes are near identical in all cases.

\subsection{Spreading drop}
\label{sec:spreading_drop}

In this second test case, we study a simple drop spreading over a no-slip substrate. We impose an equilibrium angle which governs the final shape of the drop at equilibrium. Contrary to the previous case, we need to relax the impermeability condition by allowing a controlled mass flux through the computational boundary. This introduced vertical sinking velocity, denoted as $v_\mathrm{s}$, is linked to the net mass change, being either a loss (in the case of $\theta_\mathrm{e} < \theta_0$) or a gain (in the case of $\theta_\mathrm{e} > \theta_0$) of mass\change{, with $\theta_0$ the initial angle of the drop.} The boundary condition governing the normal component of velocity at the wall is now
\begin{equation}\label{eq:toysink}
\left.u_y\right\vert_{y = \delta} = v_\mathrm{s}.
\end{equation}
Details on the numerical implementation of $v_\mathrm{s}$ are provided in Appendix~\ref{appendixA}.
\begin{figure}
\centering
\subfloat[$b = 0.109$]{
\centering
\includegraphics[width=0.3\linewidth]{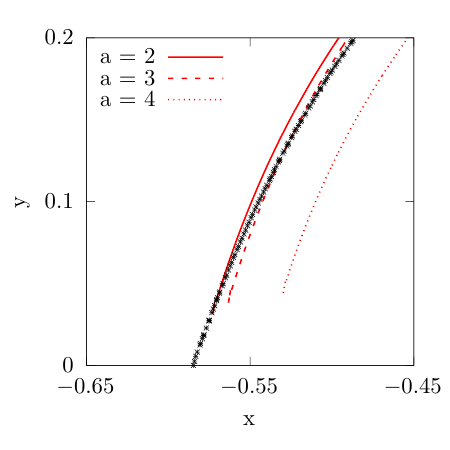}
}
\subfloat[$b = 0.217$]{
\centering
\includegraphics[width=0.3\linewidth]{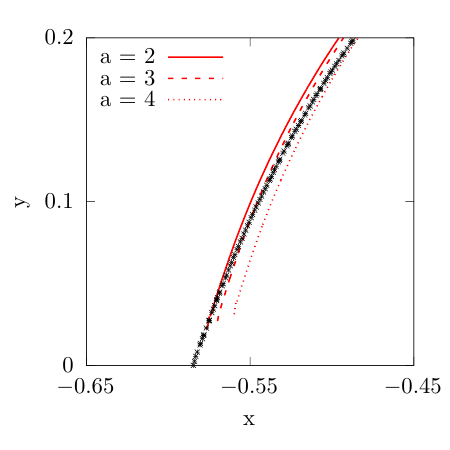}
}
\subfloat[$b = 0.434$]{
\centering
\includegraphics[width=0.3\linewidth]{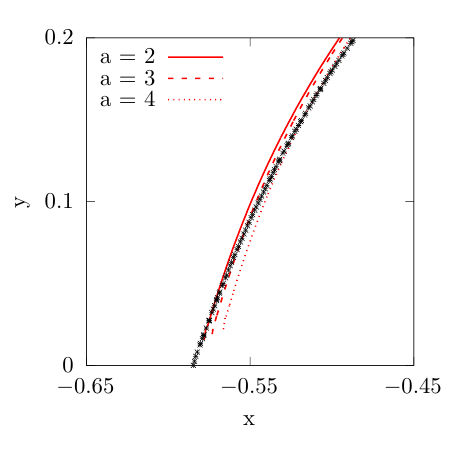}
}

\subfloat[$a = 2$]{
\centering
\includegraphics[width=0.3\linewidth]{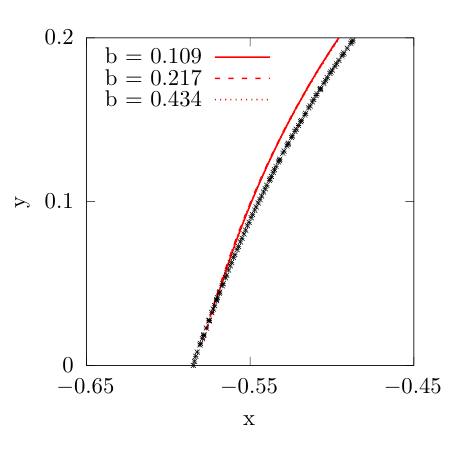}
}
\subfloat[$a = 3$]{
\centering
\includegraphics[width=0.3\linewidth]{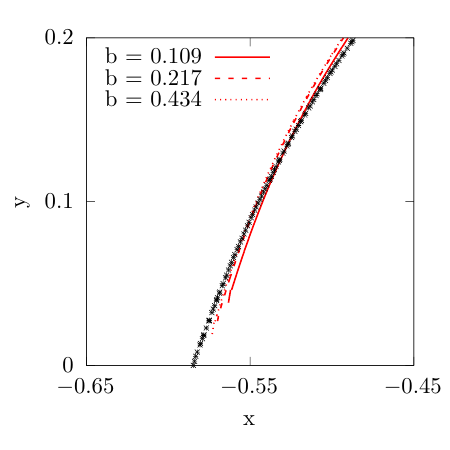}
}
\subfloat[$a = 4$]{
\centering
\includegraphics[width=0.3\linewidth]{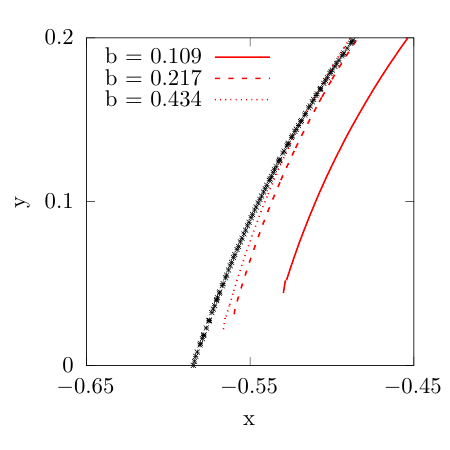}
}
\caption{Comparison of interface shapes at t = 0.1 for phase-field simulations (black lines) and volume-of-fluid toy model (red lines). (a)-(c): Effect of the variation of the geometrical factor $a$ for a fixed value of $b$. (d)-(e) Effect of the variation of $b$ for a fixed value of $a$.}
    \label{fig:a_b_variable}
\end{figure}

The initial radius of the drop is $R_0 = 0.5$ with an initial contact angle $\theta_0 = 90\degree$. The equilibrium angle is set to $\theta_\mathrm{e} =  70\degree$. The dimensionless parameters governing the flow are $\Ca =  0.0212$, $\Re =  3.978$, $\Cn =  0.01$ and $\Pe =  1$.
Fig.~\ref{fig:a_b_variable} shows the influence of varying the parameters $a$ and $b$ for this transient case. Each panel shows a close up of the left contact line region, with PF results for three different times, and the corresponding VOF results, using a range of values for $a$ and $b$. The best match is found for $a = 3$ and $b = 0.2170$ confirming the values chosen in the previous case. \change{Note that the difference in $a$ between both cases is attributed to the different equilibrium contact angle.}

\begin{figure}
\centering
\includegraphics[width=0.6\linewidth]{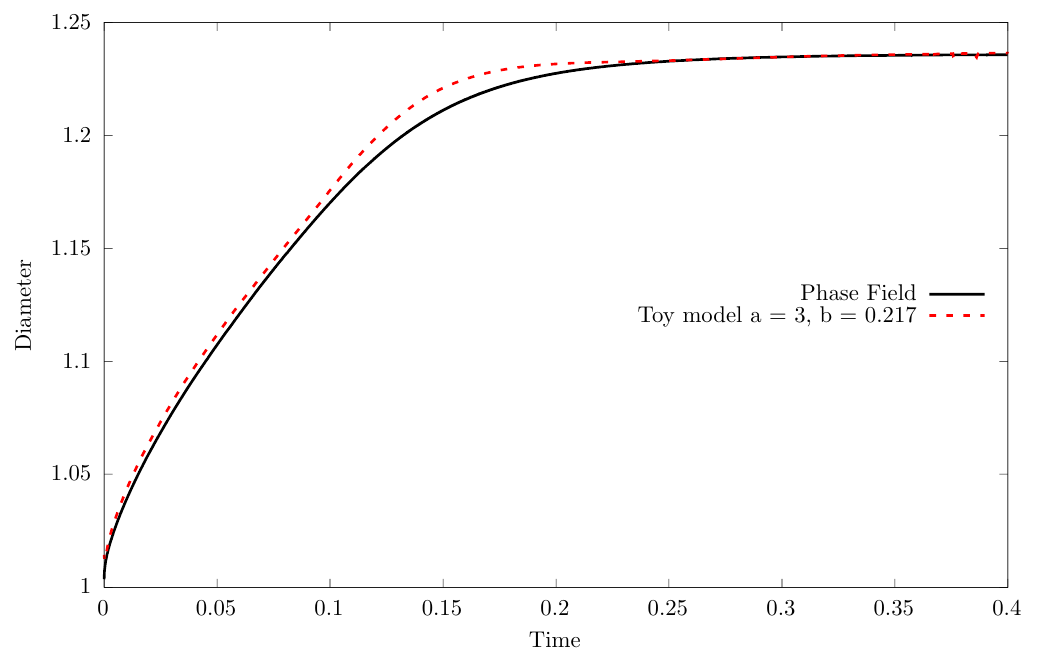}
\caption[$\:$ Spreading drop diameter comparison of the PF and toy model.]{Dimensionless diameter as a function of time for the phase-field and toy model volume-of-fluid simulations for the best values of $a$ and $b$. }
    \label{fig:toydispl}
\end{figure}
Fig.~\ref{fig:toydispl} shows the diameter of the drop as a function of time for the PF and the VOF simulations. In order to compare the displacement, the position of the contact line in the VOF setup is projected back to the wall $y = 0$ using the extracted apparent angle located at $y = \delta + \Delta$, in order to avoid the \change{numerical} difference \change{between imposed and extracted} angles $\Delta \theta_\mathrm{num}$ at $y = \delta$. From the comparison between both models, we can observe that the displacement curves superimpose almost perfectly. The final equilibrium position of the contact line is the same in both cases, implying that the sinking velocity $v_\mathrm{sink}$ allowed us to accurately represent the mass flux through the computational boundary. In the transient state, however, we see a slight difference between times $0.1$ and $0.3$ where the VOF simulation overshoots the solution. This behavior is similar to ones observed previously \change{in slip models~\cite{fullana_simulation_2022}.}

\begin{figure}
\centering
\includegraphics[width=1\linewidth]{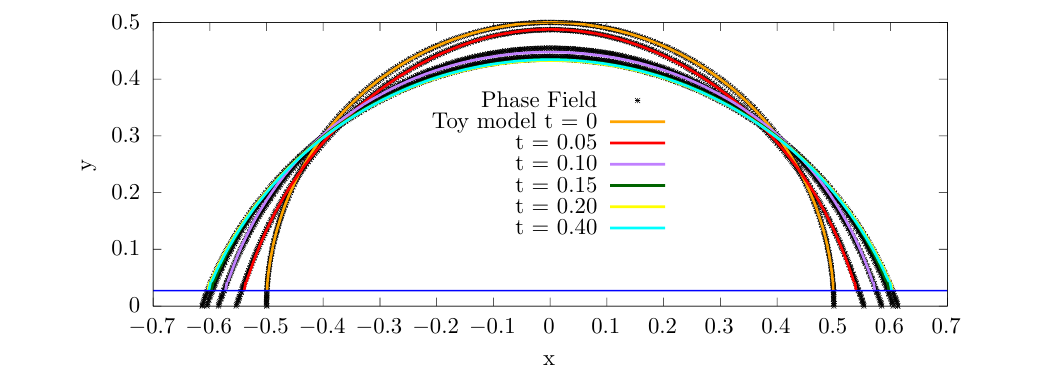}
\caption[$\:$ Spreading drop interfaces of the PF and toy model.]{Phase-field and toy model volume-of-fluid interfaces at times $t=0,\;0.05,\;0.1,\;0.15,\;0.2,\; 0.4$. The \change{blue} line corresponds to the location of the computational boundary in the toy model.}
    \label{fig:toyinterfaces}
\end{figure}
\begin{figure}
\centering
\includegraphics[width=0.6\linewidth]{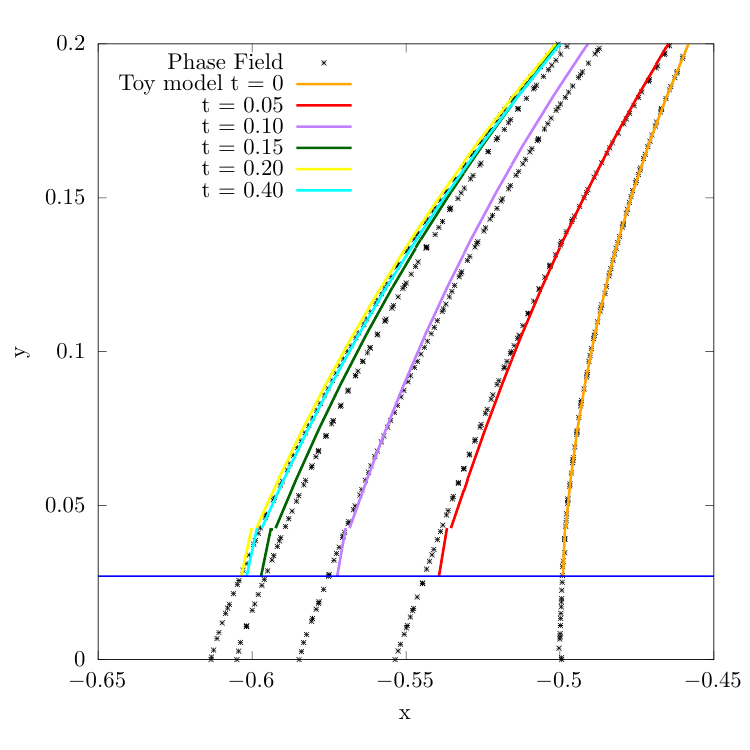}
\caption[$\:$ Zoomed interfaces of the PF and toy model.]{Zoom around the contact line of the phase-field and toy model volume-of-fluid interfaces at times $t=0,\;0.05,\;0.1,\;0.15,\;0.2,\; 0.4$. The \change{blue} line corresponds to the location of the computational boundary in the toy model.}
    \label{fig:toyinterfaces_zoom}
\end{figure}
\change{Fig.~\ref{fig:toyinterfaces} shows both the PF and VOF interfaces at $t=0,\;0.05,\;0.1,\;0.15,\;0.2,\; 0.4$, and Fig.~\ref{fig:toyinterfaces_zoom} shows a zoom of the same interfaces near the contact line. The solid blue line indicates the VOF computational boundary: as in the previous case, the domain is truncated at $y=\delta$, so the region $0 <y<\delta$ is not simulated. The superposition of the PF and VOF interfaces demonstrates overall good agreement in position and shape. A closer look in Fig.~\ref{fig:toyinterfaces_zoom} reveals a discontinuity in the final VOF segment close to the contact line. This feature manifests the numerical discrepancy $\Delta\theta_{\mathrm{num}}$, the difference between the contact angle actually reconstructed by the VOF procedure and the angle imposed via the height functions in the ghost layer.}

\section{Conclusion}
\label{sec:conclusion}
The phase-field model for dynamic wetting simulations has the advantage that it can be explicitly connected to a particular physical situation, i.e. that of a mixture of two near immiscible fluids, and thus in principle only depends on physically meaningful parameters. Computationally it is however quite expensive, and it is not suited for large scale CFD applications. On the other hand, the volume-of-fluid method which is the method of choice for such simulations, depends on ad hoc or numerical parameters for contact line motion. 

In this paper we have attempted to bridge this gap by making a local analysis of the phase-field equations near the contact line, establishing the connection between a local strong curvature and the contact line speed. VOF simulations are generally successful at capturing the interface in the bulk of the liquid, and it is only in the region near the contact line that the special features of the PF are important. Using this local analysis we have parameterized the PF results, and put them in the form of a boundary condition that can be used in VOF simulations of moving contact lines. We thus propose that this will allow VOF simulations to give results that are equivalent in many respects to the results of a full PF simulation, but at the cost of a VOF simulation. \change{We note in Appendix~\ref{appendixC} that the common case of a single component phase-change fluid can be treated in a similar way, and that an equivalent model for the contact angle can be derived, see equation~\eqref{eq:compeq_appC_toyangle}.}

\section*{Data availability} The data that support the findings of this article are openly available~\cite{fullana_2025_16927267}.

\section*{Acknowledgements}

S.Z. and T.F. thank the FLOW laboratory of KTH for hosting them and Shervin Bagheri for funding the visits in 2019, which allowed this collaboration to start. S.Z. was funded by the European Research Council (ERC) through the European Union's Horizon 2020 Research and Innovation Programme (Grant Agreement No. 883849 TRUFLOW).

\appendix
\section{Sink velocity} 
\label{appendixA}

In the toy model, a computational boundary is introduced at a distance $+ \delta$ from the wall where we apply a Navier boundary condition. When considering a spreading drop, the impermeability condition ($u_y|_{y=\delta} = 0$) needs to be relaxed in order to take into account the mass flux through that computational boundary. To that end, we introduce a sink velocity $v_\mathrm{sink}$, such that
$$ 
u_y|_{y=\delta} = v_\mathrm{sink}.
$$
To calculate the amount of mass lost during one time step, we compute the area of the polygon formed by the two contact line points, at times $t^{n-1}$ and $t^n$, and their linear projection onto the actual wall (at $y = 0$). In practice, the angle used for the projection is an apparent angle located one cell above the contact line. The computed area is then translated to the sink velocity by the simple formula
$$ 
v_\mathrm{sink} = \dfrac{1}{\tau} \dfrac{\mathcal{A}}{L},
$$
where $\tau$ is the time step, $L$ the length of the wall and $\mathcal{A}$ the signed area -- defined later. Algorithm~\ref{alg:sink} summarizes the required steps at a given instant.

\begin{figure}
\centering
\includegraphics[width=0.7\textwidth]{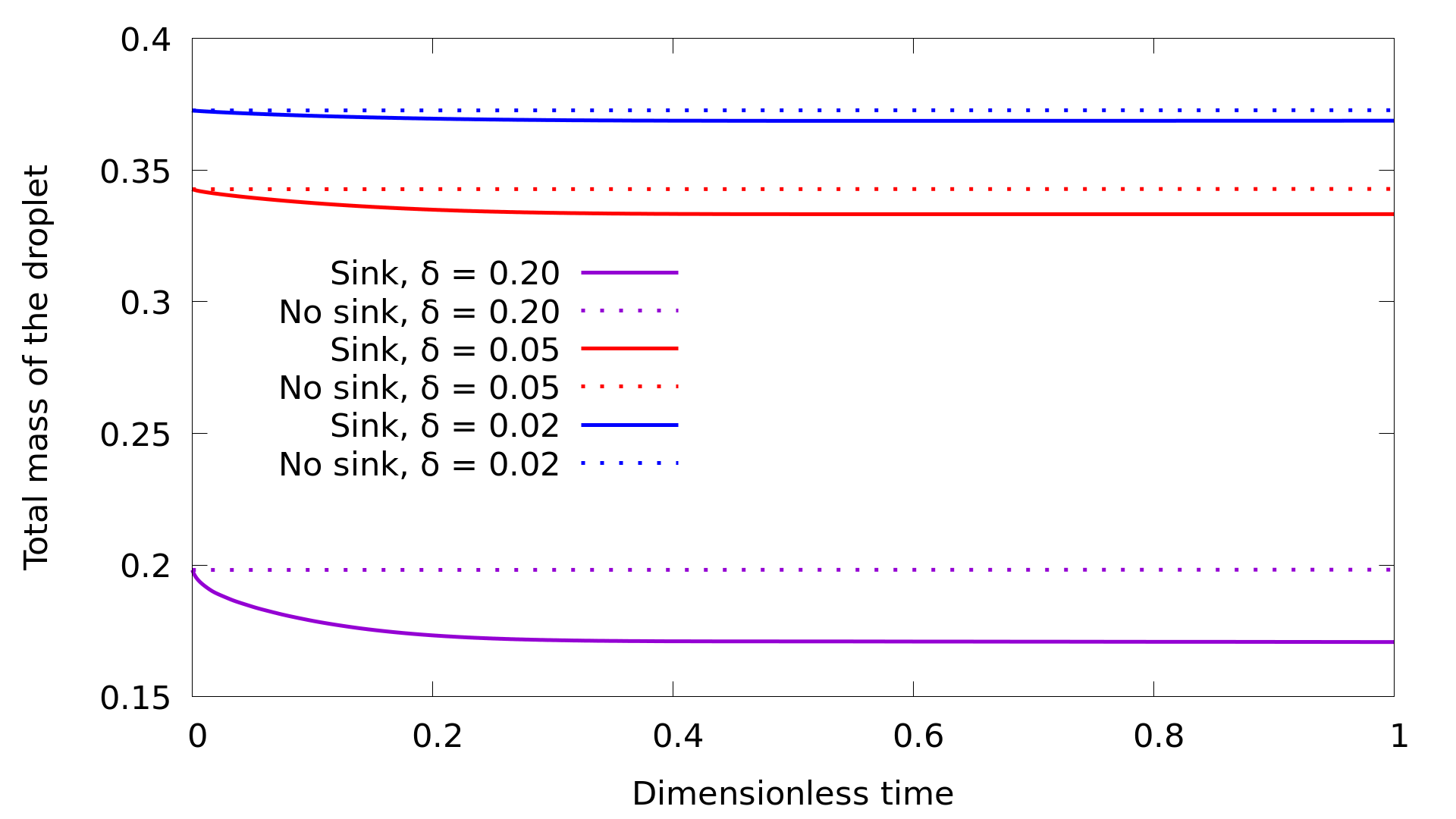}
\caption[Total mass of the drop with or without sink velocity]{Total mass of the drop as a function of time with or without sink velocity for different computational boundaries placed at $+\delta$ with $\delta = 0.02, \: 0.05, \: 0.2$.}
\label{fig:toymass}
\end{figure}

\begin{figure}
\centering
\subfloat[$\delta = 0.02$]{
\centering
\includegraphics[width=0.45\linewidth]{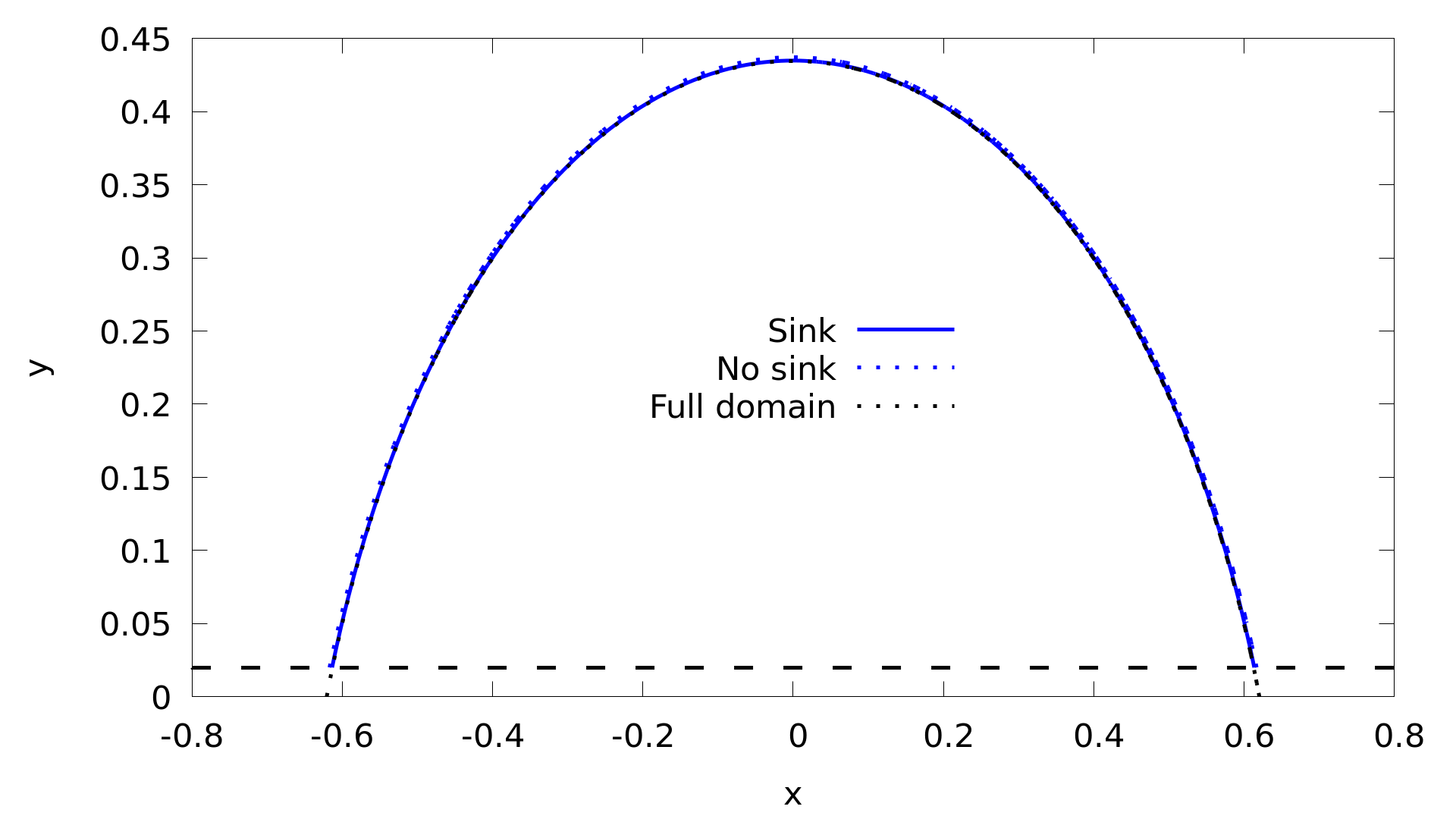}
}

\subfloat[$\delta = 0.05$]{
\centering
\includegraphics[width=0.45\linewidth]{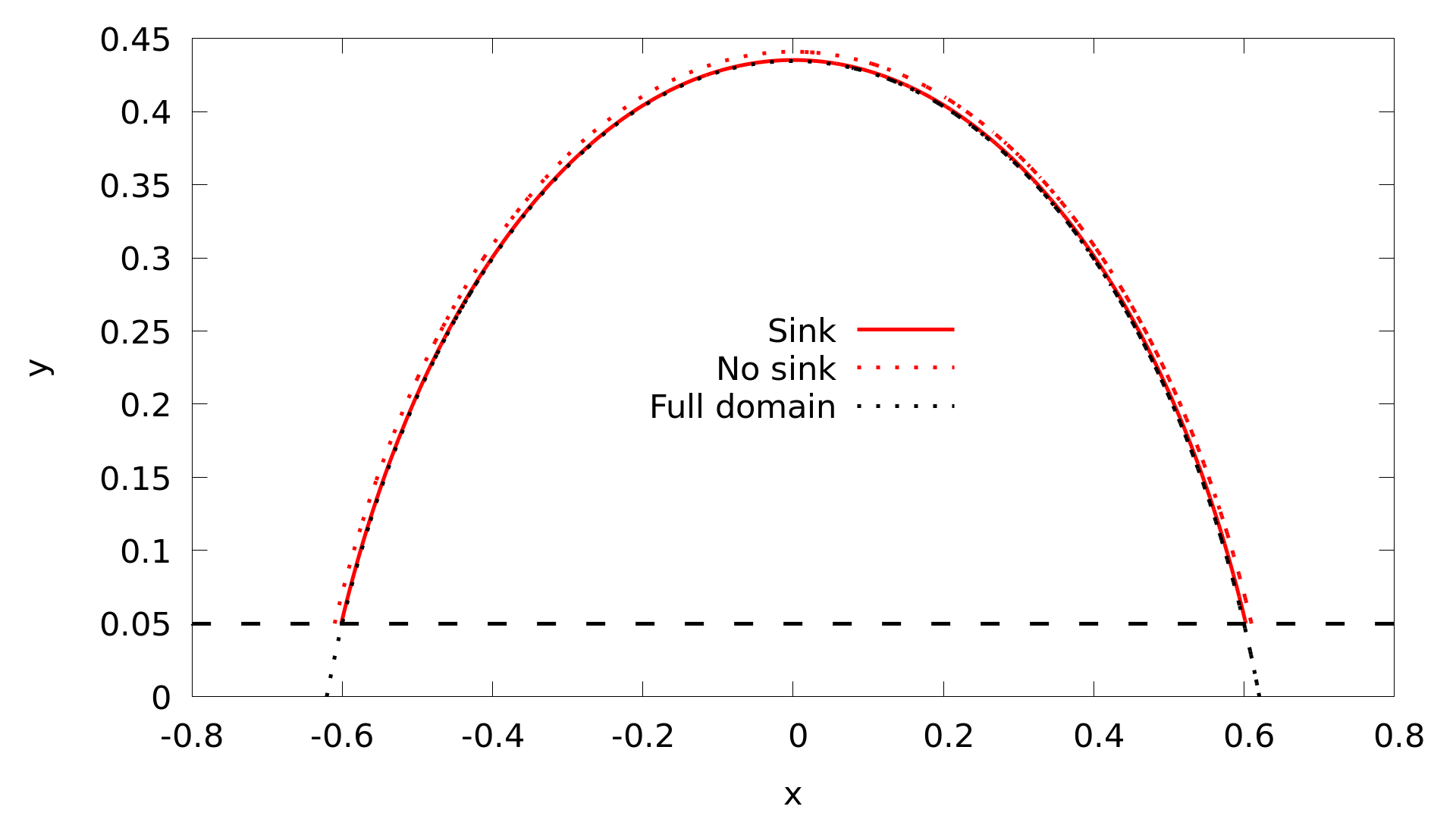}
}
\subfloat[$\delta = 0.2$]{
\centering
\includegraphics[width=0.45\linewidth]{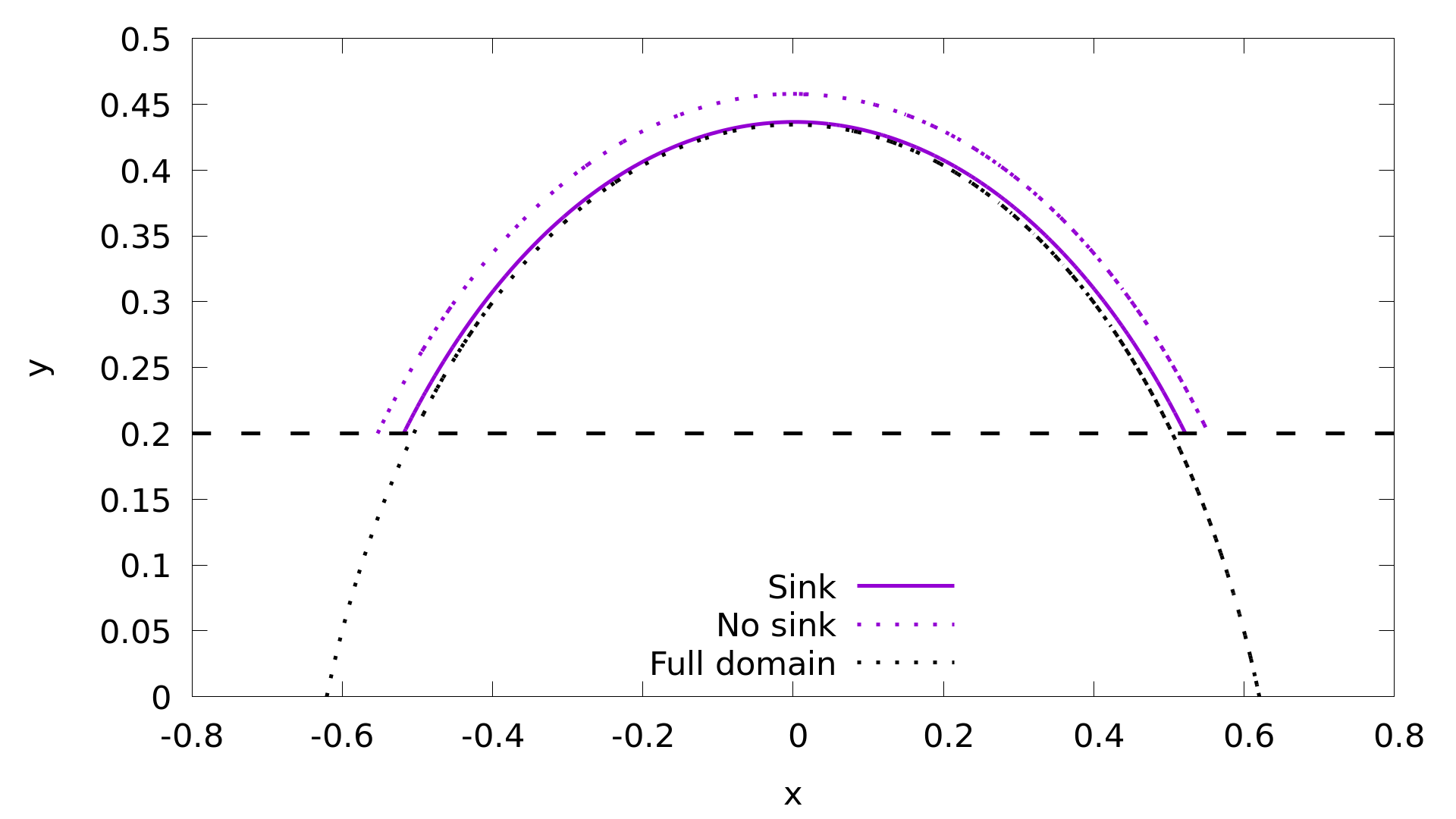}
}
    \caption[Comparison of the interfaces for different computational boundaries.]{Validation of the sink velocity imposed at the computational boundary placed at $+\delta$ with $\delta = 0.02, \: 0.05, \: 0.2$ denoted by the dotted black lines in the figures. The equilibrium angle is $\theta_\mathrm{e} = 70\degree$ for the three cases considered.}
    \label{fig:toymass2}
\end{figure}

We validate the method by considering a drop spreading over a substrate for different values of $\delta$. The drop is initialized with a diameter $D = 1$ in a $2 \times 2$ domain and we apply a constant contact angle $\theta_\mathrm{e} = 70\degree$. The viscosities, densities and the surface tension are set to 1. The simulations are run until the equilibrium position is reached. In Fig.~\ref{fig:toymass}, we show the total mass as a function of time when applying the sink velocity for $\delta = 0.2, \: 0.05, \: 0.02$. The initial mass inside the drop decreases as spreading occurs. In Fig.~\ref{fig:toymass2}, we compare the final interfaces for three cases: (i) toy model with sink velocity (ii) toy model without sink velocity (iii) full domain. From the results, we can observe the sink velocity allows us to match to the final true equilibrium shape (full domain). The error in final interface is a function of $\delta$. As $\delta$ is increased, the linear extrapolation procedure -- that does not take into account the curved interface at the contact line -- induces a higher difference in final shapes. In practice, when using the toy model on a real setup, $\delta$ is small compared to the domain size.
\begin{algorithm*}
    \SetKwData{Left}{left}
	\SetKwData{This}{this}
	\SetKwData{Up}{up}
	\SetKwInOut{Input}{input}\SetKwInOut{Output}{output}

	\For{each cell}{
	Locate the cell one grid point above the contact line\\
	Compute the contact line position $x^n_\mathrm{CL}$ at $y = \delta + \Delta$\\
	Compute the apparent angle $\theta_\mathrm{app}$ using the unit normal $\mathbf{n}$\\
	Project the contact line position $$\tilde{x}^n_\mathrm{CL} = x^n_\mathrm{CL} + (\delta + \Delta) \: \tan \left(\dfrac{\pi}{2} - \theta_\mathrm{app} \right)$$\\
	Compute the polygon area $\mathcal{A}$ formed by the points $$(x^n_\mathrm{CL}, \delta + \Delta), \: (\tilde{x}^n_\mathrm{CL}, 0), \: (\tilde{x}^{n-1}_\mathrm{CL}, 0), \: (x^{n-1}_\mathrm{CL}, \delta + \Delta)$$\\
	where $x^{n-1}_\mathrm{CL}$ and $\tilde{x}^{n-1}_\mathrm{CL}$ were stored in the previous time step\\
	Determine the sink velocity
	$$ v_\mathrm{sink} = \dfrac{1}{\tau} \dfrac{\mathcal{A}}{L} $$\\
	Store $x^{n-1}_\mathrm{CL} = x^n_\mathrm{CL}$ and $\tilde{x}^{n-1}_\mathrm{CL} = \tilde{x}^n_\mathrm{CL}$
	}
	\caption{Sink velocity pseudo-code}\label{alg:sink}
\end{algorithm*}

\newpage 
{\color{black}\section{Toy model calculation of parameter $b$} 

\label{appendixB}


Rescaling the steady version of~\eqref{eq:diffueq} using a length-scale $\ell$ and a magnitude $\phi \sim \Pe \ell$ according to~\eqref{eq:delta_toyDiff}, we get

\begin{equation} \label{eq:diffueq_appB}
 \mathbf{u} \cdot \tilde{\nabla} C  = \tilde{\nabla}^2 \tilde{\phi},
\end{equation}

where $\tilde{\nabla}=\ell \nabla$ and $\tilde{\phi}=\phi / (\Pe \ell) $. This equation is thus universal in the sense that all dependencies of $\Ca, \Pe, \Cn$ etc are scaled out and equation \ref{eq:diffueq_appB} could be solved once and for all. The only remaining independent input parameter is the equilibrium contact angle $\theta_e$. The correspondingly rescaled version of the momentum equation will introduce also the ratio of the viscosities in the two fluid regions.   

In order to arrive at a simple quantitative model, the dimensionless order unity parameter $b$ is introduced in equation~\eqref{eq:delta_toyDiff_eq}. This parameter is expected to essentially capture the equivalent of solving eq \ref{eq:diffueq_appB}, and we thus expect that it may depend on $\theta_e$ and the viscosity ratio, but that it is independent of  $\Ca, \Pe, \Cn$ etc.

As a bold attempt at estimating $b$ the following model equation was analyzed:

\begin{equation} \label{eq:diffueq_appB_2b}
\Pe \delta \dfrac{d C}{d x}   = \dfrac{d^2 }{dx^2} ( - \dfrac{d^2 C}{dx^2} -\frac{Cn}{R} \dfrac{dC}{dx}+\Psi'(C) ),
\end{equation}

Similarly to the treatment in~\eqref{eq:pot_toy}--\eqref{eq:chemconditions}, we consider a perturbation solution around the equilibrium solution in~\eqref{eq:zerothsoln}. The equation for the first order perturbation becomes: 

\begin{equation} \label{eq:diffueq_appB_1st_order}
-\frac{\Pe \delta}{\Cn} \dfrac{dC_0}{dx}  + \frac{1}{R} \dfrac{d^3C_0}{dx^3} = \dfrac{d^2 }{dx^2} \left( - \dfrac{d^2 C_1}{dx^2} +\Psi''(C_0)  C_1
\right),
\end{equation}

Multiplying both sides by an $F(x)$ such that $\dfrac{dF(x)}{dx} = C_0(x)$, and repeatedly integrating the right hand side partially, it is seen that the right hand side vanishes, yielding a solvability condition on the remaining terms:

\begin{equation} 
\label{eq:diffueq_appB_solvability}
\frac{\Pe \delta}{\Cn} \int  F \dfrac{d^2 F}{d x^2}  dx  = \frac{1}{R} \int  F \dfrac{d^4 F}{dx^4} dx
\end{equation}

Rearranging this, using~\eqref{eq:delta_toyDiff_eq}, as
\begin{equation} 
\label{eq:diffueq_appB_solvability2}
\displaystyle \frac{\Pe \delta}{\Cn / R} = \frac{1} {2b} \frac{1}{2} \frac{2 \sqrt{2}}{3}  = \dfrac{\int  F \dfrac{d^4 F}{dx^4} dx}{\int  F \dfrac{d^2 F}{d x^2}  dx}   
\end{equation}

The integrals on the right hand side can both be calculated, giving $\int  F \dfrac{d^4 F}{dx^4} dx = \dfrac{2 \sqrt{2}}{3}$ and $\int  F \dfrac{d^2 F}{d x^2}  dx=2 \sqrt{2} (1-\ln 2)$ .

This yields

\begin{equation} 
\label{eq:diffueq_appB_b_equals}
b = \frac{1-\ln 2}{\sqrt{2}}  = 0.2170
\end{equation}

It should be emphasized that the derivation of this number is by no means a rigorous proof, and that $b$, as well as the parameter $a$ introduced in~\eqref{eq:toyNS}, are to be adjusted according to comparisons between the model in~\eqref{eq:Rab}--\eqref{eq:toyangle}, and full solutions of the phase field equations, as shown in Fig.~\ref{fig:a_b_variable}. As seen there however, the value $b=0.2170$ gives satisfactory results, and it has been used throughout.}

{\color{black}\section{Single component fluid with phase change.} 

\label{appendixC}


In this appendix we sketch a similar development for the case of a single component fluid with phase change, i.e. a volatile droplet in its own vapor. The van der Waals equation of state would be a first example.

Instead of~\eqref{eq:cahn-hil-dim} and~\eqref{eq:momPF}, we will need to begin from the full compressible equations, including the energy equation

\begin{equation} 
\label{eq:compeq_appC_mass}
\dfrac{\partial \rho}{\partial t} + (\rho u_i),_{i} = 0,
\end{equation}

\begin{equation}
\label{eq:compeq_appC_mom}
\dfrac{\partial \rho u_i}{\partial t} + (\rho u_i  u_j)_{,j} = -P_{ij,j} + (\mu (u_{i,j}+u_{j,i}))_{,j},
\end{equation}

\begin{equation}
\label{eq:compeq_appC_en}
\dfrac{\partial \rho e}{\partial t} + (\rho e u_j)_{,j} = -P_{jk} u_{k,j} + \mu (u_{k,j}+u_{j,k}) u_{k,j} +  (\lambda T_{,j})_{,j},
\end{equation}

\begin{equation}
\label{eq:compeq_appC_Press}
P_{ij} = \left(p(\rho,T) - \kappa \left(\rho \rho_{,kk} + \frac{1}{2} \rho_{,k} \rho_{,k} -\frac{\rho}{T} \rho_{,k} T_{,k} \right) \right)\delta_{ij} +  \kappa \rho_{,i} \rho_{,j}.
\end{equation}

In addition to density, velocity and viscosity $\rho$, $u_i$ and $\mu$, we have internal energy $e$, temperature $T$ and heat conductivity $\lambda$, $\mu$ and $\lambda$ taken constant for simplicity. $P_{ij}$ denotes the pressure tensor, including the Korteweg stress \cite{anderson_AnnRevFluMech1998}, 
and $\kappa$ is the gradient energy coefficient. The function $p(\rho,T)$ is the equation of state.

The equililbrium solution to these equations, corresponding to~\eqref{eq:zerothsoln}, is found for $u_i=0$ and $T=const$, and $P_{ij}=p_s(T) \delta_{ij}$, where $p_s(T)$ is the liquid-vapor coexistence pressure at temperature $T$. In implicit form this can be written as

\begin{equation}
\label{eq:compeq_appC_eqsoln}
f(\rho,T) + p_s(T) = \frac{\kappa}{2} \rho_x^2,
\end{equation}

where the free energy $f$ is related to the pressure as $p=-f+\rho f_\rho$ (indices denoting derivatives).

Analogous to \eqref{eq:zerothorder0}--\eqref{eq:chempot_interface}, we now examine the departure from this equilibrium due to interface curvature, and thus derive the Gibbs-Thomson relation for this specific diffuse interface model with a gradient energy. We make a change of variable as before, $r=R+\epsilon x$ where $\epsilon <<1$ indicates that curvature is much larger than the interface width. Introducing perturbed quantities as

\begin{equation}
\label{eq:appC_pert}
\begin{aligned}
P{ij}&=p_s(T) \delta{ij}+\epsilon P_{ij}^1, \\
\rho&=\rho_0 + \epsilon \rho_1, \\
u_i&=0 + \epsilon u_i^1, \\
e&=e_0 + \epsilon e_1, \\
T&=T_s + \epsilon T_1 .\\
\end{aligned}
\end{equation}

Introducing these in $P_{11}$ and expanding we find

\begin{equation}
\label{eq:compeq_appC_expansion}
\begin{aligned}
P_{11}&=P_{11}^0 + \epsilon P_{11}^1, \\
P_{11}^0 &= p_s(T_s), \\
P_{11}^1&= T_1 \dfrac{dp_s}{dT} + \kappa (\rho_1 \rho_0 f_{\rho \rho} +\rho_{0x} \rho_{1x} - \rho_{1} \rho_{0xx}  -\rho_{0} \rho_{1xx}) + \kappa \frac{1}{R} \rho_0 \rho_{0x}. \\
\end{aligned}
\end{equation}

Multiplying $P_{11}^1$ by $\rho_{0x} / \rho_0^2$ and integrating we can obtain a solvability condition similar to~\eqref{eq:chempot_interface}. The second term in $P_{11}^1$ can be shown to vanish using the relation in~\eqref{eq:compeq_appC_eqsoln}.
The remaining terms are then 

\begin{equation}
\label{eq:compeq_appC_solvability}
\begin{aligned}
  T_1 \dfrac{dp_s}{dT} \int_{v}^{l} \frac{\rho_{0x}}{\rho_0^2} dx  + \kappa \frac{1}{R} \int_{v}^{l} \frac{\rho_{0x}^2}{\rho_0}dx &=0,\\
  \text{where}\\
  \int_{v}^{l}  \frac{\rho_{0x}}{\rho_0^2} dx &= \frac{1}{\rho_v}-\frac{1}{\rho_l}= \frac{1}{\tilde{\rho}},\\
  \kappa \int_{v}^{l} \frac{\rho_{0x}^2}{\rho_0}dx &= \int_{v}^{l} \frac{\sqrt{2\kappa(f+p_s)}}{\rho_0} d\rho = \frac{\sigma}{\overline{\rho}}.\\
\end{aligned}
\end{equation}

Here we denote the density corresponding to difference in specific volume $\tilde{\rho}$ as $\tilde{\rho}=1/(1/\rho_v-1/\rho_l)$. In the last integral the equilibrium solution in~\eqref{eq:compeq_appC_eqsoln} was used, we identified the surface tension as $\sigma=\int_{v}^{l}\sqrt{2\kappa(f+p_s)} d\rho$, and introduced an averaged density as

\begin{equation}
\label{eq:compeq_appC_avdensity}
\overline{\rho}=\frac{\int_{v}^{l}\sqrt{2\kappa(f+p_s)}  d\rho}{\int_{v}^{l} \frac{\sqrt{2\kappa(f+p_s)}}{\rho_0} d\rho}.
\end{equation}

The Clapeyron equation in this context reads 
\begin{equation}
\label{eq:compeq_appC_Clapeyron}
 \dfrac{dp_s}{dT} =\frac{L}{T (1/\rho_v-1/\rho_l)}= \frac{L \tilde{\rho}}{T}.
\end{equation}
Here $L$ denotes the latent heat.
The first equation in~\eqref{eq:compeq_appC_solvability} can then be written as
\begin{equation}
\label{eq:compeq_appC_GT}
  T_1 \dfrac{L}{T}  + \frac{1}{R} \frac{\sigma}{\overline{\rho}} = 0, 
\end{equation}

i.e. the Gibbs-Thomson relation. We note the explicit expression for the average density in~\eqref{eq:compeq_appC_avdensity}, and that this would emphasize the low density of the gas phase.

To proceed, we now derive a scaling for the region in the vicinity of the contact line, with a length scale $\delta$, to be determined. The velocity scale $u$ is the liquid velocity. The lowest order temperature is the equilibrium temperature $T_s$, and $T_1$ is the perturbation of the temperature. In analogy with~\eqref{eq:delta_toyDiff_eq} we assume that the main balance in the energy equation~\eqref{eq:compeq_appC_mom} is between the thermal diffusion $(\lambda T_,j)_{,j}$ and the $pdV$ like term $P_{jk}u_{k,j}$, which at first order is $p_s u_{k,k} \sim p_s  (\rho_l/\rho_v)(u/\delta) $. Depending on the fluid properties it is also possible that the thermal diffusion is balanced instead by the convective term on the lefthand side of \eqref{eq:compeq_appC_en}, whch would give slightly different estimates.   As in~\eqref{eq:delta_toyDiff_eq} we introduce a nondimensional order unity factor $b$, which is to be calibrated by comparison between the model and full simulations
\begin{equation}
\label{eq:compeq_appC_en_scaling}
p_s  \frac{\rho_l}{\rho_v}\frac{b u}{\delta} = \lambda \frac{T_1}{\delta^2}.
\end{equation}
In the momentum equation C2 we take the balance between the viscous term and the surface tension, estimating $P_{11}$ as  $P_{11} \sim T_1 (dp_s/dT) $ and using~\eqref{eq:compeq_appC_Clapeyron}. As in~\eqref{eq:toyNS} we introduce a nondimensional order unity parameter $a$
\begin{equation}
\label{eq:compeq_appC_NS_scaling}
 T_1 \frac{L \tilde{\rho}}{T_s} \frac{1}{\delta} = \mu \frac{a u}{\delta^2}.
\end{equation}
Here $1/\tilde{\rho} =(1/\rho_v-1/\rho_l)$.

Equations \eqref{eq:compeq_appC_GT}, \eqref{eq:compeq_appC_en_scaling} and \eqref{eq:compeq_appC_NS_scaling} are now the counterparts to equations \eqref{eq:chempot_interface}, \eqref{eq:delta_toyDiff_eq} and \eqref{eq:toyNS}, and are to be solved for $\delta$, $R$ and $T_1$. The result of this, corresponding to equations \eqref{eq:Rab}, \eqref{eq:delta_ab} and \eqref{eq:chempot_interface}, are
\begin{equation}
\label{eq:compeq_appC_R}
 R =-\frac{1}{a} \frac{\delta}{(\mu u / \sigma)} \frac{\tilde{\rho}}{\overline{\rho}}= - \frac{1}{\sqrt{ab}}\frac{1}{(\mu u /\sigma)} \frac{\tilde{\rho}}{\overline{\rho}}\sqrt{\frac{\rho_v}{\rho_l}} \sqrt{\frac{\lambda \mu}{\tilde{\rho} p_s (L/T_s)}},
\end{equation}
\begin{equation}
\label{eq:compeq_appC_delta}
 \delta =\sqrt{\frac{a}{b}} \sqrt{\frac{\rho_v}{\rho_l}} \sqrt{\frac{\lambda \mu}{p_s\tilde{\rho}(L/T_s)} },
\end{equation}
\begin{equation}
\label{eq:compeq_appC_T1}
 T_1 = u \sqrt{a b} \sqrt{\frac{p_s \mu}{\lambda (L/T_s)} \frac{\rho_l}{\rho_v \tilde{\rho}}}.
\end{equation}
We can now write down the model for the apparent contact angle, as in~\eqref{eq:toyangle}, which is seen to be very similar:
\begin{equation}
\label{eq:compeq_appC_toyangle}
\cos(\theta_\mathrm{a}) = \cos(\theta_\mathrm{e}) + \dfrac{\delta}{R}  = \cos(\theta_\mathrm{e}) + a \frac{\tilde{\rho}}{\overline{\rho}} \Ca,
\end{equation} 
with the capillary number $\Ca=\mu u / \sigma$.

}

\bibliography{biblist}

\end{document}